\documentclass[manuscript]{aastex}
\usepackage{graphics}
\usepackage{graphicx}

\begin{document}

\title{Transient Mass Loss Analysis of Solar Observations using Stellar Methods}

\author{M. K. Crosley}
\affil{Johns Hopkins University, Department of Physics \& Astronomy, 3400 N. Charles Street, Baltimore, MD 21218}

\author{R. A. Osten}
\affil{Space Telescope Science Institute, 3700 San Martin Dr, Baltimore, MD 21218}
\affil{Center for Astrophysical Sciences, Johns Hopkins University, Baltimore, MD 21218}

\and

\author{C. Norman}
\affil{Johns Hopkins University, Department of Physics \& Astronomy, 3400 N. Charles Street, Baltimore, MD 21218}

\begin{abstract}

Low frequency dynamic spectra of radio bursts from nearby stars offer the best chance to directly detect the stellar signature of transient mass loss on low mass stars.
\citet{Crosley2016} proposes a multi-wavelength methodology to determine coronal mass ejection parameters, such as Coronal Mass Ejection (CME) speed, mass, and kinetic energy.

We test the validity and accuracy of the results derived from the methodology by using Geostationary Operational Environmental Satellite X-ray observations and Bruny Island Radio Spectrometer radio observations.
These are analogous observations to those which would be found in the stellar studies.
Derived results from these observations are compared to direct white light measurements of the Large Angle and Spectrometric Coronagraph.

We find that, when a pre-event temperature can be determined, that the accuracy of CME speeds are within a few hundred km/s, and are reliable when specific criteria has been met.  
CME mass and kinetic energies are only useful in determining approximate order of magnitude measurements when considering the large errors associated to them.  
These results will be directly applicable to interpretation of any detected stellar events and derivation of stellar CME properties.

\end{abstract}


\section{Introduction}  
			
Just as coronal mass ejections (CMEs) from the sun are an important component of space weather on Earth, they are also important to consider for planets around other cool stars.  
Stellar flares are routinely observed on cool stars but clear signatures of stellar CMEs have been less forthcoming. 
Lack of experimental evidence creates a dependance on solar scaling relations to estimate the impact of these events with no measure of the validity of their use.

Traditional solar observations of CMEs use coronagraphs to observe Thomson scattering of photospheric photons off coronal electrons \citep{2006ApJ...642.1216V}.
This emission can be used to obtain parameters such as angular size, height-time relations, mass, and rate of occurrence.   
This white light emission of a CME is faint when compared to the integrated solar disk emission necessitating the use of a coronograph.   
Current astronomical coronagraphs cannot achieve sufficient star contrast to detect a CME at a distance of 1 - 2 R$_\ast$ from the star \citep{Mawet2012}, making it infeasible for CME observations.  
Therefore, only integrated stellar disk emission can be used for observations. 

The relationships between solar and stellar flares may provide a useful tool to analyze stellar CMEs.
These relationships have been studied and continue to be expanded upon.  
A few observed stellar counterparts to solar flare phenomena include white light flares, nonthermal gyrosynchrotron emission, X-ray emission, FUV emission, and coherent radio emission \citep{Osten2016}. 
Although the details are still unclear, this multi-wavelength nature of the flare process suggests that both solar and stellar flares share the same basic physical processes. 
The Neupert effect, an observational feature of many flares whereby the time-integrated microwave flux closely matches the X-Ray emission temporal behavior, has also been observed in both the solar and stellar cases \citep{Gudel1996}

There is reason to be skeptical despite the appearance of a strong connection between solar and stellar flares.
Detailed comparisons of solar and M dwarf flares show a difference in the nature of accelerated particles. 
M dwarf flares produce more accelerated particles due to a perpetuated hot plasma and a quiescent nonthermal microwave component during times when no obvious flares are occurring \citet{Gudel1996}.
Also, solar particles do not penetrate the photosphere deeply enough to reproduce the observed magnitude of the white light stellar flare signal \citep{Kowalski2015} causing difficulties in the application of the standard solar flare model to M dwarfs \citep{2006ApJ...644..484A}.

There exists solar flare-solar CME relationships in addition to a strong solar-stellar flare relationship.
\citet{2006ApJ...650L.143Y} has shown using Geostationary Operational Environmental Satellite (GOES) X-ray measurements that solar CMEs have an increasing association rate to solar flares as flare flux, flare peak flux, and flare duration increase untill saturating at 100\%. 
\citet{Aarnio2012} and \citet{Drake2013} have both found an empirical relationship between the solar flare X-ray energy and its associated CME mass.
Physically driven arguments show that there exists a rough equipartition between the total radiated energy of the flare and the mechanical energy of its associated CME \citep{Emslie2005,Osten2015}. 

A connection between stellar flares and stellar CMEs is supported by relationship between solar flares and solar CMEs in conjunction with the strong relationship between solar and stellar flares.
\citet{Crosley2016} proposed a multi-wavelength analysis of radio emissions stemming from the CME and flare light curves which provides a means to measure CME properties (mass, velocity, and occurrence rate).
Since stellar observations are unable to make use of current astronomical coronagraphs, only full disk integrated light may be utilized for CME detections.  
\citet{Crosley2016} discusses the use of optical or X-ray flare energy observations in tandem with radio observations of type II burst as the best way of detecting and constraining CME properties.  

A type II burst is a non-thermal radio emission originating from a fast mode magnetohydrodynamic (MHD) shock \citep{2006GMS...165..207G}.
The shock is generated by the passage of a CME through the stellar atmosphere at sufficient speeds.  
They appear as a slowly drifting radio burst following an exponential path through frequency with time related to the speed of the shock (and thus the speed of the CME) and the ambient density of the local corona \citep{2006GMS...165..207G}. 
The shocks accelerate non-thermal electrons, which in turn produce radio emission at the fundamental and harmonic of the local plasma frequency via well-known plasma processes \citep{2006GMS...165..207G}. 
The velocity of the shock can be constrained from analysis of the dynamic spectra given constraints on the coronal characteristics.  
The shock velocity can be used to constrain the velocity of the CME generating the shock.  

Several questions arise from the stellar astronomer's perspective.  
This paper aims to answer some unresolved questions: how well do the physical parameters returned from the dynamic spectra match directly observed values and how robust are the solar scaling relationships?  
Do the masses returned from empirical relationships agree with those derived from energy equipartition relationships and to observed values?
How well can observed CME kinetic energy be reproduced though combinations of CME mass and velocity or through scaled flare energy relationships?

In Section 2 reviews the methodology for interpreting stellar flare and stellar CME data.  
Section 3 reviews the choices of solar data which would best represent stellar observations.  
The analysis procedure is presented in Section 4.  
The results and discussion are presented in Section 5.
Section 6 concludes.

\section{Methodology for Interpreting Stellar CME-Flare Data}

The following sections recreate the multi-wavelength analysis proposed in \citet{Crosley2016}.

\subsection{Low Frequency Radio Observations}

A type II burst is an easily identifiably signal unique to CMEs.
A MHD shock will be produced when the CME is traveling with sufficient velocity through the stellar atmosphere.
Langmuir waves are generated by electrons which are accelerated by the shock and radiate via the local plasma frequency and its harmonics.
This frequency changes as the source travels through the atmosphere producing a distinct slope in time and frequency.
The frequency will vary in time as:
\begin{equation}
\frac{d\nu}{dt}=\frac{\partial\nu}{\partial n}\frac{\partial n}{\partial h}\frac{\partial h}{\partial s}\frac{\partial s}{\partial t} 
\end{equation}
where $\nu$ is frequency, $n$ is electron density, $h$ is radial height above the star, $s$ is distance along the path which the shock travels, and $t$ is time.  

The drift rate $\left(\frac{d\nu}{dt}\right)$ is thus composed of four terms describing the changing environment around the shock.  
As the shock propagates away or towards the star, the local electron density $(n)$ will change.
The shocks emits via the local plasma frequency which depends on the local density as:  $\nu_{p} = \sqrt{\frac{ne^{2}}{\epsilon_{0}m_{e} } } $ in SI, where $m_{e}$ is the electron mass, $e$ is the electron charge, and $\epsilon_{0}$ is the permittivity of free space. 
An outward moving shock will emit at a lower frequencies as it travels towards lower densities.

The second term $\left(\frac{\partial n}{\partial h}\right)$ describes how the density changes as a function of height radially above the stellar corona. 
A barometric model $\left(\frac{\partial n}{\partial h} = \frac{- n}{H_{0}} \right)$ of the stellar atmosphere is used where $H_{0}$ describes the density scale height. 
It is known however, from sources such as \citet{Leblanc1999} and \citet{Guhathakurta1999}, that the radial density profile of the Sun is not a simple barometric model.
In the stellar case, we do not have similar constraints on the density profiles other than the barometric constraint.

Table 1 shows the expected emitted frequency as a function of radial distance from the star.
It compares the expected frequencies from a Barometric density model, the \citet{Leblanc1999} density model, and the \citet{Guhathakurta1999} coronal hole density.
The Barometric model tends to decay slower at all stages and thus will return higher plasma frequencies than the other two solar models for the same distance. 

\begin{deluxetable}{c | c c c c c c c c c}	
\tabletypesize{\footnotesize}
\tablecaption{Frequency Model Comparison}
\tablecolumns{8}
\tablewidth{0pt}
\tablehead{
\vspace{ -1.21cm}  
}
\startdata
	Distance in R$_{\odot}$ & 1 & 1.1 & 1.2 & 1.3 & 1.4 & 1.5 & 1.6 & 1.7 & 1.8 \\ \cline{2-10}
	Barometric [MHz] & 119.3 & 100.5 & 84.6 & 71.3 & 60.0 & 50.6 & 42.6 & 35.9 & 30.2\\
	\citet{Guhathakurta1999} [MHz] & 119.3 & 62.6 & 35.2 & 21.1 & 13.6 & 9.5 & 7.1 & 5.6 & 4.7 \\
	\citet{Leblanc1999} [MHz] & 82.7 & 62.5 & 48.5 & 38.4 & 31.0 & 25.4 & 21.2 & 17.8 & 15.2\\
	\enddata
\tablecomments{\citet{Leblanc1999} models average destiny at solar minimum as : $n_{e} = 3.3\times10^{5} r^{-2} + 4.1\times10^{6} r^{-4.09} + 8.0\times10^{7} r^{-6} $ cm$^{-3}$ where it was normalized to $n_{e}(1au) = 7.2$cm$^{-3}$.  
\citet{Guhathakurta1999} models the coronal hole density as $n_{e} = (1736.9 r^{-13.72} + 19.95 r^{-4} + 1.316 r^{-2})\times 10^{5} $ cm$^{-3}$.
The units of r are in R$_{\odot}$.
The Guhathakurta model is used to set the density for the barometric model.  
}
\end{deluxetable}
 
The final two terms describe the path and speed the CME takes as it travels. 
The CME does not necessarily travel perpendicularly outward and so $\frac{\partial h}{\partial s} = cos\theta$ describes how the vertical height changes as a function of the path traveled.
Here $\theta$ is the angle at which the shock is traveling relative to the radial direction.
The distances examined are sufficiently small such that any acceleration of the CME is negligible and will therefore the CME will have a constant speed $\left(\frac{ds}{dt} = v_{s}\right)$.   

Substituting all the differential terms leads to the drift rate expression:
\begin{equation}
\label{eq:nudot}
\frac{d\nu}{dt} = \left(\frac{\nu}{2n}\right)\left(-\frac{n}{H_{0}}\right)\left(cos\theta\right)\left(v_{s}\right) = -\frac{\nu v_{s} cos\theta} {2H_{0}}  
\end{equation}
where $\nu$ is frequency, $v_s$ is velocity of the shock, $\theta$ is the angle at which the shock path is traveling relative to the radial direction, $H_{0}$ is the density scale height.

The emitted frequency ($\nu$) of the type II burst is directly measured in observations.  
Measuring how the observed frequency changes with time produces the drift rate.
The density scale height ($H_{0}$) of the corona is modeled from the coronal temperature via:
$H_{0} = \frac{k_{B}T}{\mu m_{p}g} = 5.01 T\times 10^{3}$ cm/K where $k_{B}$ is the Boltzmann constant, $m_{p}$ is the proton mass, $\mu$ is the mean molecular weight of the particles equal to 0.6, and g is the local gravity of the star which depends on its mass and radius.  
X-ray measurements are used to constrain the coronal temperature closer to the star.
Integrated light measurements are proportional to density squared which is biased towards over dense regions.

Assuming a perpendicular path $\left( cos\theta \approx 1\right)$, the only parameter left unconstrained is $v_{s}$.
A detection of a drifting radio burst provides a lower limit constraint on the speed of the CME shock and therefore is a lower limit on the speed of the CME itself.   
Additionally, the uniqueness of type II bursts to CMEs allows a measurement of type II burst occurrence rate to be a lower bound for the occurrence rate of CMEs as not every CME produces a type II burst.

\subsection{Flare Energy Observations}

\citet{2006ApJ...650L.143Y} has shown, using GOES X-ray measurements, that solar CMEs have an increasing association rate to solar flares as flare flux, flare peak flux, and flare duration increase.
The association rate saturates to 100\% when the flare are sufficiently large.  

Building from this connection, \citet{Osten2015} has shown evidence supporting the existence of an equipartition between the total radiated energy of the flare and the mechanical energy of its associated CME. 
This relation can be described as:  
\begin{equation}
\frac{1}{2} M_{CME} v^{2} = \frac{E_{rad}}{\epsilon f_{rad}} 
\end{equation}
where $f_{rad}$ is the fraction of the bolometric radiated flare energy appropriate for the waveband in which the energy of the flare is being measured \citep{Osten2015}.
The factor $\epsilon$ is a constant of proportionality $\approx 0.3$ to describe the relationship between bolometric radiated energy and CME kinetic energy.  
The velocity of the CME is $v$, its total mass is $M_{CME}$, and radiated energy is $E_{rad}$.

The GOES (1-8 \AA) bandpass is utilized in this paper and thus the flare observations ($E_{flare}$) produce:
\begin{equation}
\label{eq:equipar}
\frac{1}{2} M_{CME} v^{2} = \frac{E_{GOES}}{\epsilon f_{GOES}} 
\end{equation}
where $f_{GOES} = 0.06$ is the fraction of the bolometric flare energy designated to the GOES bandpass of 1-8 \AA \ \citep{Osten2015}. 

\citet{Aarnio2012} and \citet{Drake2013} have found an empirical relationship between the solar flare X-ray energy and its associated CME mass.
It is a relationship of the form:
\begin{equation}
\label{eq:Mcme}
M_{CME} = A E^{ \gamma} \mbox{  [g]}
\end{equation}
where $M_{CME}$ is the CME mass, A is a constant of proportionality, and $\gamma$ describes the power law.  
Both papers have a similar functional form but flare energy is defined uniquely for each.  
The flare energy for \citet{Aarnio2012} (E$_{triangular/tri.}$) is defined as one half of the observed peak flux of the GOES (1-8 \AA) band, times the observed flare duration, times 4$\pi(1$ AU$)^{2}$. \citet{Drake2013} defines the flare energy (E$_{integrated/int.}$) as the X-ray fluence in the GOES (1-8 \AA) band.  
For \citet{Aarnio2012}, $A = 2.7 \pm 1.2 \times 10^{-3}$ in cgs units and $\gamma = 0.63 \pm 0.04$.
For \citet{Drake2013} these parameters are: $A = 10^{-1.5 \mp 0.5} $ in cgs units and $\gamma = 0.59 \pm 0.02$.   

\citet{Drake2013} also derived a relationship for the kinetic energy (KE) of the CME described in the same manner as equation (\ref{eq:Mcme}).
It takes the form:
\begin{equation}
\label{eq:KEcme}
KE_{CME} = B E^{\beta} \mbox{  [erg]}
\end{equation}
, where $KE_{CME}$ is the CME kinetic energy, $B = 10^{0.81 \mp 0.85}$ in cgs units and $\beta = 1.05 \pm 0.03$ \citep{Drake2013}.

These two formulas (eqn. (\ref{eq:equipar}) and (\ref{eq:Mcme})) provide two methods for determining CME mass through the use of flare measurements (CME velocity determined from radio measurements). 
Assuming that both provide similar values for $M_{CME}$, they can be utilized in concert to solve for either the flare energy as a function of CME velocity:
\begin{equation}
\label{eq:Combined}
E_{GOES} = \left[ \frac{A\epsilon v^{2}}{2} f_{GOES}  \right]^{\frac{1}{1-\gamma}}
\end{equation}
or the velocity as a function of flare energy:
\begin{equation}
\label{eq:VCombined}
v = \sqrt{\frac{2}{A\epsilon f_{GOES}}}(E_{GOES})^{\frac{1-\gamma}{2}}
\end{equation}
.
These are now the single variable equations $E_{GOES}(v)$ and $v(E_{GOES})$.  

\section{Solar Observations}

Solar observations are used to test the accuracy and validity of the formulas derived in the previous section to examine their application for stellar observations.  
To best emulate stellar observations, the GOES X-ray satellite is used for full disk observations of the Sun to determine solar x-ray temperatures and flare properties while the Bruny Island Radio Spectrometer (BIRS) is used for radio observations of solar dynamic spectra. 
The direct white light observations of solar CMES detected by the Large Angle and Spectrometric Coronagraph (LASCO) on board The Solar and Heliospheric Observatory (SOHO) are compared to the derived CME properties of velocity, mass, and kinetic energy.

The criteria used for data selection:
\begin{enumerate}
	\item An observed flare had to be M or X-class, that is the flare had peak flux above $10^{-5}$ or $10^{-4}$ W/m$^{2}$ 
	\item The flare had an associated CME observed with the white light coronograph
	\item The CME had a type II burst associated to it.  
\end{enumerate}

In general, the sun has many CMEs, type II bursts, and flares.  
However, our goal is to best examine the types of events that would be found on M dwarf stars.  
Criterion 1 selects only the largest, and thus most infrequent, events that are closet to the types of flares found on M dwarfs.  
The number of events again decreases with criterion 2, and then again with criterion 3 leaving only a handful of events that fulfill all requirements.  
We took this small pool of fully comprehensive data to test the validity of this approach before committing the time and resources into developing a larger pool of multi-wavelength data.  
  
The LASCO CME events were identified between 2-6 R$_\odot$.  
The BIRS observations were typically between 6-62 MHz.  
Using the \citet{Guhathakurta1999} coronal hole density equation used in Table 1, this corresponds to distances between $\sim$1.1 to $\sim$ 1.7R$_{\odot}$. 
For the barometric model, this would represent inferred distances between $\sim$1.4 to $\sim$ 2.8R$_{\odot}$.

\subsection{Solar Coronagraph Observations of CMEs}

The catalog\footnote{\label{note1}http://cdaw.gsfc.nasa.gov/CME\_list/index.html} 
of all CME's manually identified since 1996 from the LASCO/SOHO is used to identify CME's and their physical properties for analysis (see \citet{CmeCatalog2009} for more information).  
The CME's linear speed and mass are retrieved from the catalog while its kinetic energy can be determined using these two values.

There are three different markings on the GOES Flare Class designation for events in the tables presented in this paper.  
These are: $\ast$, $\dagger$, and $\dagger'$.  
The $\ast$ indicates that a CME remarked as a Poor Event in the SOHO/LASCO catalog.  
Dynamic spectra that do not follow an exponential shape are marked by a $\dagger$.  
Finally, there were some dynamic spectra which had multiple events.  
The $\dagger'$ indicates that an event had two events; one with an exponential shape and one with a non-exponential shape.
The exponential shaped event was chosen for analysis in these type of events.   

Table 2 summarizes the recordings for events that fulfilled the selection criteria above.
The LASCO/SOHO white light coronagraph C2 observes the field-of-view from 2 to 6 solar radii.     
In the case of a Halo CME (a CME directed towards earth), these measurements become more difficult  necessitating the use of models for height-time measurements \citep{Yashiro2004} and mass measurements \citep{Vourlidas2000}  

 \begin{deluxetable}{l c | cccccc}
	\tabletypesize{\footnotesize}
	\tablecaption{SOHO/LASCO Measurements\tablenotemark{[1]}}
	\tablewidth{0pt}
	\tablehead{\vspace{ -1.21cm}}  
	\startdata 
	GOES & Date & First C2 & Angular Width & Linear Speed & Acceleration & Mass & Kinetic Energy \\
	Class & of Event & Appearance [UT] & [deg] & [km/s] & [m/s$^{2}$]  &  [10$^{15}$ g] & [10$^{30}$ erg]  \\    \cline{1-8}
	X\tablenotemark{\dagger'} & 2012-07-06 & 23:24:06 & 360 & 1828 & -56.1 & 8.4\tablenotemark{[3]} & 140\tablenotemark{[3]}  \\
	X & 2013-11-08 & 03:24:07 & 360 & 497 & 11.7 & 8.8\tablenotemark{[3]} & 11\tablenotemark{[3]}     \\
	X & 2014-04-25 & 00:48:03 & 296 & 456 & -9.1 & 4.6\tablenotemark{[3]} & 4.8\tablenotemark{[3]}   \\  \cline{1-8}
	M & 2011-01-28 & 01:25:46 & 119 & 606 & -20.9 & 3.5 & 6.4 \\
	M\tablenotemark{*}\tablenotemark{\dagger} & 2011-02-13 & 18:36:05 & 276 & 373 & 24.4\tablenotemark{[2]} & 0.96\tablenotemark{[3]} & 0.67\tablenotemark{[3]} \\ 
	M\tablenotemark{*} & 2012-03-17 & 22:12:05 & 64 & 66 & 4\tablenotemark{[2]} & 0.79 & 0.017  \\
	M\tablenotemark{\dagger} & 2012-06-03 & 18:12:05 & 180 & 605 & -8.7 & 3.1\tablenotemark{[3]} & 5.6\tablenotemark{[3]}  \\
	M & 2013-05-02 & 05:24:05 & 99 & 671 & 1.1 & 3.3 & 7.4\\
	M\tablenotemark{*} & 2013-10-24 & 01:25:29 & 360 & 399 & -17\tablenotemark{[2]} & 1.9\tablenotemark{[3]} & 1.5\tablenotemark{[3]}\\
	M\tablenotemark{*} & 2014-01-08 & 04:12:05 & 294 & 643 & -3.7\tablenotemark{[2]} & 2.5 & 5.1  \\
	M\tablenotemark{*} & 2014-02-20 & 00:12:05 & 50 & 271 & -1\tablenotemark{[2]} & 0.96 & 0.35  \\
	M & 2014-03-20 & 04:36:06 & 360 & 740 & -2\tablenotemark{[2]} & 4.5\tablenotemark{[3]} & 12\tablenotemark{[3]} \\
	M & 2014-11-03 & 23:12:33 & 155 & 638 & -23.9 & 2.3\tablenotemark{[3]} & 4.7\tablenotemark{[3]}   \\
	M & 2014-12-17 & 05:00:05 & 360 &  587 & -2.1 & 12\tablenotemark{[3]} & 20\tablenotemark{[3]} \\ 
	\enddata	
	\tablecomments{
	Summary of the SOHO/LASCO measurements for events that fulfilled selection criteria. 
	The C2 white light coronagraph images from 2 to 6 solar radii.
		}
\tablenotetext{[1]}{
Data taken from http://cdaw.gsfc.nasa.gov/CME\_list/index.html
}
\tablenotetext{[2]}{
Acceleration is uncertain due to either poor height measurement or a small number of height-time measurements (See Section 3.4 of \citet{Yashiro2004} for details).
}
\tablenotetext{[3]}{
Mass and Kinetic Energy are uncertain due to poor mass estimate. Mass measurement assumes that the CME material is in the sky-plane. Kinetic Energy is obtained from the mass and linear speed (see \citet{Vourlidas2000} for details).  
}
\tablenotetext{*}{
CME remarked as a Poor Event in the SOHO/LASCO CME catalog
}
\tablenotetext{\dagger}{
Dynamic Spectra did not follow an exponential shape.
}
\tablenotetext{\dagger'}{
Dynamic Spectra had two events, one with an exponential shape and one with a non-exponential shape.  The exponentialy shaped event was chosen for analysis. 
}
\end{deluxetable}

\subsection{GOES Flare Light Curves}

GOES data\footnote{\label{note0}satdat.ngdc.noaa.gov/sem/goes/data/new\_avg/} was used for X-ray observations drawing from the GOES15 satellite data when available.
The X-ray flux is measured in two bandpasses: short (S = 0.5 - 4 \AA) and long (L = 1 - 8 \AA).
Flare energies are measured using the long bandpass flux over the period of the flare. 
The Sun's temperature is measured by examining the ratio (S/L) of these bandpasses \citep{Thomas1984}.
This temperature is used to determine the density scale height ($H_{0}$) required to model the type II burst from the BIRS dynamic spectra.

Observations utilized a 1-minute cadence in both the long and short bands.
These observations have no spatial resolution and are full disk measurements of the Sun.
This is representative of stellar observations which observe the full stellar disk even if the wavelength coverage is higher energy than typical astronomical X-ray missions.

\subsection{Low Frequency Solar Radio Observations}

The BIRS catalog\footnote{\label{note2}www.astro.umd.edu/~white/gb/index} was used to identify and constrain radio bursts associated with CME's for analysis.  
The dynamic spectra is typically recorded between 6 - 62 MHz at a 3 minute cadence. 
These observations also have no spatial resolution and are full disk measurements of the Sun.
This is representative of stellar observations which observe the full stellar disk.

\begin{figure}
\centering
\includegraphics[angle=-90,scale=.5,clip=true, trim=0cm 0cm 0cm 0cm]{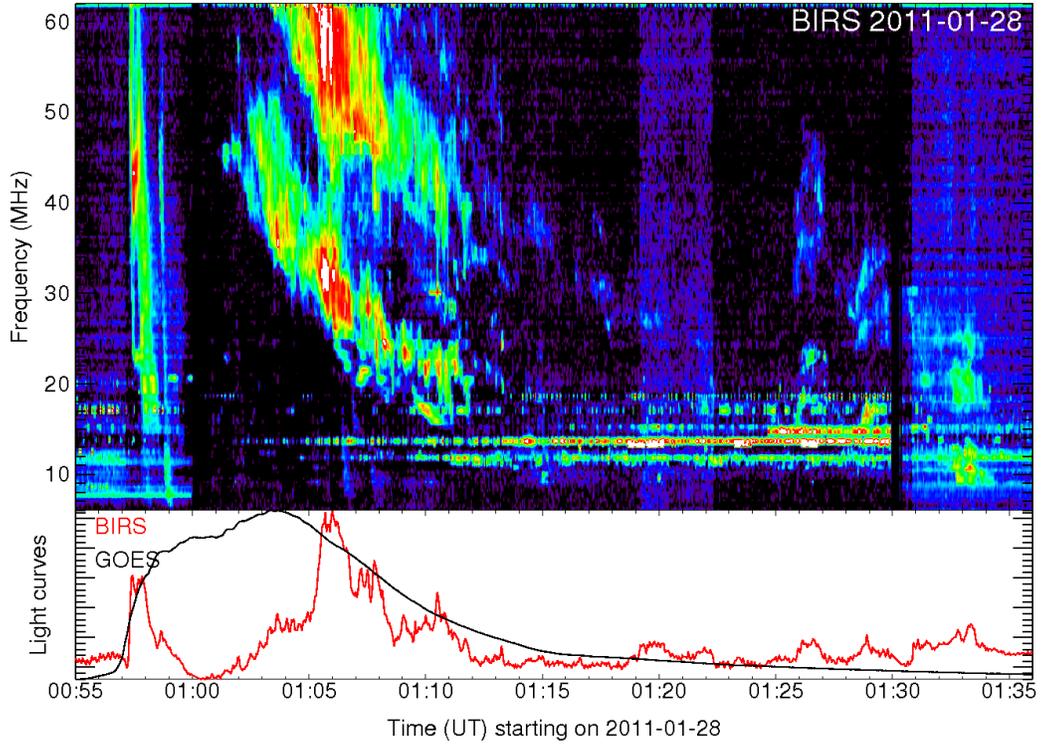} 
\caption{
Radio dynamic spectra and GOES light curve for M-class eruptive event occurring on 2011-01-28.
The top panel displays the BIRS radio dynamic spectra where a more red color denotes a more intense signal and a blue color is a fainter signal.
The bottom panel displays the GOES light curve in black.
It also displays a trace of a higher frequency line in the BIRS spectra range in red.  
}
\label{fig:f1}
\end{figure}

Figure \ref{fig:f1} shows an example of the BIRS dynamic spectra containing a type II burst associated to a M class GOES X-ray flare.
The vertical axis is frequency in MHz and the horizontal axis is time (UT) in mins.  
The type II bursts has both the first and second harmonics of the plasma frequency.  
This allows for both harmonics to be used to determine drift rates for velocity calculations. 
There is  also some radio frequency interference in the low frequencies near the ionospheric cutoff of $\sim$10 MHz.
Two curves are presented below the dynamic spectra.
These are the GOES light curve in black and a single frequency within the dynamic spectra is plotted in red.  
This is to show the timing of the burst relative to the flare.   
The remaining images are shown in the appendix.

\section{Analysis}

\subsection{X-Ray Light Curves}
The GOES X-ray data provides information regarding the flare and the solar corona via its temperature.  
The flares' peak flux, fluence, and duration in the GOES long passband are recorded to measure the flare properties for that event.  
Two methods for measuring flare energy are used.
The first follows \citet{Aarnio2012} and defines the flare total flux ($E_{tri.}$) as one half of the observed peak flux, times the observed flare duration, times $4\pi (1$AU$^{2}$).
The second follows \citet{Drake2013} and defines the flare total flux ($E_{int.}$) as the integrated flux over the observed flare duration, times $4\pi (1$AU$^{2}$).

The ratio (R = S/L) of the GOES short passband (S) to the GOES long passband (L) is used to model the coronal temperature.
\citet{Thomas1984} created an analytical curve, which is fit to the ratio of detector responses as a function of plasma temperature.  
The simple fit for temperature is $T(R) = 3.15 + 77.2R - 164R^{2} + 205R^{3}$ [$10^{6}$ K].
The fit was modeled on temperatures between $4 \times 10^{6}$K and $30 \times 10^{6}$K.  

Observations are analyzed using two unique temperatures. 
The first approach defines each event with its own unique temperature.
A flares' pre-flare quiescent state temperature is measured and used to determine the coronal scale height for that event.
The quiescent state is defined as an approximately steady state minimum ($\sim 10^{-6}$ W/m$^{2}$) flux for a period before the impulse phase of the flare.  
The quiescent state is modeled by a constant and the error is the standard deviation of this value. 
The duration of the quiescent state chosen to model the temperature varies for each event but is typically 20-40 mins.

The second approach averages the temperatures, described in the step above, found for each event to determine a single average temperature with its own error.
The average temperature is simply an average of the values and the error is the standard deviation of the mean ($\sigma = \bar{\sigma}/\sqrt{Count}$).
This is more akin to stellar observations which may not have a time resolved temperature available during observations.   
Just as in the first approach, the average temperature is used to find a single scale height to be used for all events.  
For this particular dataset, the temperature and scale height found were: temperature T = $4.0 \pm 0.5 \times 10^{6}$ K  and scale height $H_{0} = 2.0 \pm 0.2 \times 10^{10}$ cm.  

The GOES X-ray measurements are weighted to measure near the base of the corona.  
The readings are sensitive to the emission measure, which is proportional to the density squared times emitting regions volume.  
The scale height determined from this measurement extends outward due to the nature of the barometric model.

\citet{Su2016} and \citet{Wan2016} observed the 2014-01-08 CME event where part of the analysis included using differential emission measure analysis to characterize the thermal properties of and around the emitted CME.  
It was found the surrounding corona had a temperature $\sim$2 MK in both analysis.  
The temperature using the method of \citet{Thomas1984} determines a temperature of 3.7 MK and the average coronal temperature for this set is 4 MK. 

This may be suggestive of an overestimation on behalf of the GOES temperature.  
A rough factor of 1/2 to the temperature, and thus the scale height, would be useful for the points which are either 'poor' or do not fit into the exponential model well.  
The shorter scale height also makes the barometric model more comparable to the \citet{Guhathakurta1999} density profile.
Conversely, a 1/2 factor makes the `good' points less accurate when compared to the coronagraphic velocity. 
The $\sim$ 2 MK difference between the two temperatures is within the variation found among the GOES temperature between unique events therefore no scaling factor will be included for the rest of the analysis.
  
Only the results of the average temperature case will be discussed for the remainder of this paper.  
The results of the unique temperature analysis are, in general, more accurate but qualitatively the same.  

\subsection{Radio Dynamic Spectra}

The BIRS radio dynamic spectra can now be modeled using the scale heights determined from the GOES analysis.
The dynamic spectra identification follows the methodology of \citet{Osten2006}. 
A background noise level is determined by finding the root-mean-square (RMS) deviation in a region devoid of observed signals in the dynamic spectra.  
The maxima are defined as maximum above some factor times the RMS in the region of the dynamic spectra with a type II burst.
The maxima are then grouped together through an algorithm with critical distances in time and frequency. 

 Any groups associated to the type II burst are linearly modeled when plotted as Ln(Frequency) vs. time.
The slope (m) of this line corresponds to $m  = \frac{v_{s}cos\theta}{2H_{0}}$.  
Using $cos\theta \approx 1$, the shock speed will be: $v_{s} = 2H_{0}m$ which is a lower limit to the CME speed.
There are often harmonic signals in these events which can be compared to each other to check the consistency of the determined velocity.

\subsection{Error Assumptions}

The goal of the error analysis is to determine the sources of random error and understand the relative magnitude of these errors.  
Relative differences between determined and observed values can be more important than statistical uncertainty in a derived value. 

Several asssumptions are made in regards to the tabulated errors.  
The GOES N Series Data Book\footnote{\label{note4}http://goes.gsfc.nasa.gov/text/GOES-P\_Databook.pdf} states that the GOES long band has a threshold flux of $2\times 10^{-8}$ W/m$^{2}$ with a signal-to-noise $>1$ within a 10s period.
The errors are determined for each 1 min increment.  
The error for a peak flux is just the 1 min error while the integrated errors use a combination of the errors for each step in the flares duration.  
The errors are unlisted as they are small; the peak flux error is $\sim$1.5\% and integrated flux error is $\sim0.5\%$.   

When calculating the drift rate/velocity from drift rate, cos$\theta$ is assumed to be $\approx1$ and has error $\delta_{\theta} = 0$.  
The values and errors of  $\epsilon$ and $f_{GOES}$ are not well defined so the errors are assumed to be zero ($\delta_{\epsilon} = \delta_{f} = 0$).
Any calculations that use these parameters will have systematically low errors.  

Additional information regarding error specifics is found within appendix section \ref{sec:EA}. 

\section{Results and Discussion}

Table 3 is a reference table of all values used throughout analysis. 

\begin{deluxetable}{l | l}
	\tabletypesize{\footnotesize}
	\tablecaption{List of Variables used in Analysis}
	\tablewidth{0pt}
	\tablehead{  \vspace{ -1.21cm}  }
	\startdata 
	Parameter & \hspace{5.5cm} Description \\  \cline{1-2}
	$\ast$ & CME remarked as a Poor Event in the SOHO/LASCO CME catalog  \\
	$\dagger$ & Dynamic Spectra did not follow an exponential shape. \\
	$\dagger$' & Found multiple signals, the exponential signal was used for analysis  \\ \cline{1-2}
	E$_{int.}$  & Flare energy defined as the time integration of the X-light curve, times $4\pi(1AU)^{2}$. \\
	E$_{tri.}$ & Flare energy defined as one-half of the peak flux, times the flare duration, times $4\pi(1AU)^{2}$.  \\
	E$_{DS,int.}$ & Flare energy from equation (\ref{eq:Combined}): $\left[ \frac{A\epsilon f_{GOES}}{2} V_{DS}^{2} \right]^{\frac{1}{1-\gamma}}$ using the Drake constants $\gamma = 0.59 $, $A = 10^{-1.5}$  \\ 
	E$_{DS,tri.}$  & Flare energy from equation (\ref{eq:Combined}): $\left[ \frac{A\epsilon f_{GOES}}{2} V_{DS}^{2} \right]^{\frac{1}{1-\gamma}}$ using the Aarnio constants $\gamma = 0.63 $, $A = 2.7 \times 10^{-3}$ \\  \cline{1-2}
	V$_{Cg}$ & Coronagraphic CME velocity measurement \\
	V$_{DS}$ & Velocity derived from the dynamic spectra. \\  
	V$_{E_{int.}}$ & Velocity from equation (\ref{eq:VCombined}): $\sqrt{\frac{2}{A\epsilon f_{GOES}}}(E_{int.})^{\frac{1-\gamma}{2}}$, using the Drake constants $\gamma = 0.59 $, $A = 10^{-1.5}$ \\
	V$_{E_{tri.}}$ & Velocity from equation (\ref{eq:VCombined}): $\sqrt{\frac{2}{A\epsilon f_{GOES}}}(E_{tri.})^{\frac{1-\gamma}{2}}$, using the Aarnio constants $\gamma = 0.63 $, $A = 2.7 \times 10^{-3}$  \\	\cline{1-2}
	M$_{Cg}$ & Coronagraphic CME mass measurement \\
	M$_{tri.}$ & CME mass determined using Aarnio's empirical relation: $AE_{tri.}^{\gamma}$,  $\gamma = 0.63 $, $A = 2.7 \times 10^{-3}$ \\
	M$_{int.}$ & CME mass determined using Drake's empirical relation:  $ AE_{int.}^{\gamma}$, $\gamma = 0.59 $, $A = 10^{-1.5}$ \\
	M$_{DS,tri.}$  &  CME mass determined by the equipartition equation using $V_{DS}$ and $E_{tri.}$: $M_{CME} = \frac{2E_{tri.}}{V_{DS}^{2}\epsilon f_{GOES}}$   \\
	M$_{DS,int.}$ &  CME mass determined by the equipartition equation using $V_{DS}$ and $E_{int.}$: $M_{CME} = \frac{2E_{int.}}{V_{DS}^{2}\epsilon f_{GOES}}$   \\ \cline{1-2}
	KE$_{Cg}$ & CME kinetic energy determined by coronagraphic measurements, $\frac{1}{2}M_{Cg}V_{Cg}^{2}$\\
	KE$_{int.}$ & CME kinetic energy determined using Drake's empirical relation: $BE_{int,}^{\beta}$, $B = 10^{0.81}$ , $\beta$ = 1.05   \\
	KE$_{DS,tri.}$ & Left hand side of equation (\ref{eq:equipar}) whilst using $M_{tri.}$: $\frac{1}{2}M_{tri.}V_{DS}^{2}$ \\
	KE$_{DS,int.}$ &  Left hand side of equation (\ref{eq:equipar}) whilst using $M_{int.}$: $\frac{1}{2}M_{int.}V_{DS}^{2}$. \\ 
	KE$_{E,tri.}$  &  Right hand side of equation (\ref{eq:equipar}) whilst using $M_{tri.}$: $\frac{E_{tri.}}{\epsilon f_{GOES}}$ \\
	KE$_{E,int.}$  & Right hand side of equation (\ref{eq:equipar}) whilst using $M_{int.}$: $\frac{E_{int.}}{\epsilon f_{GOES}}$ \\
	\enddata
\end{deluxetable}

\subsection{GOES Observations}

\begin{figure}
\centering
\includegraphics[angle=-90,scale=0.5,angle=0,clip=true, trim=0cm 0cm 0cm 0cm]{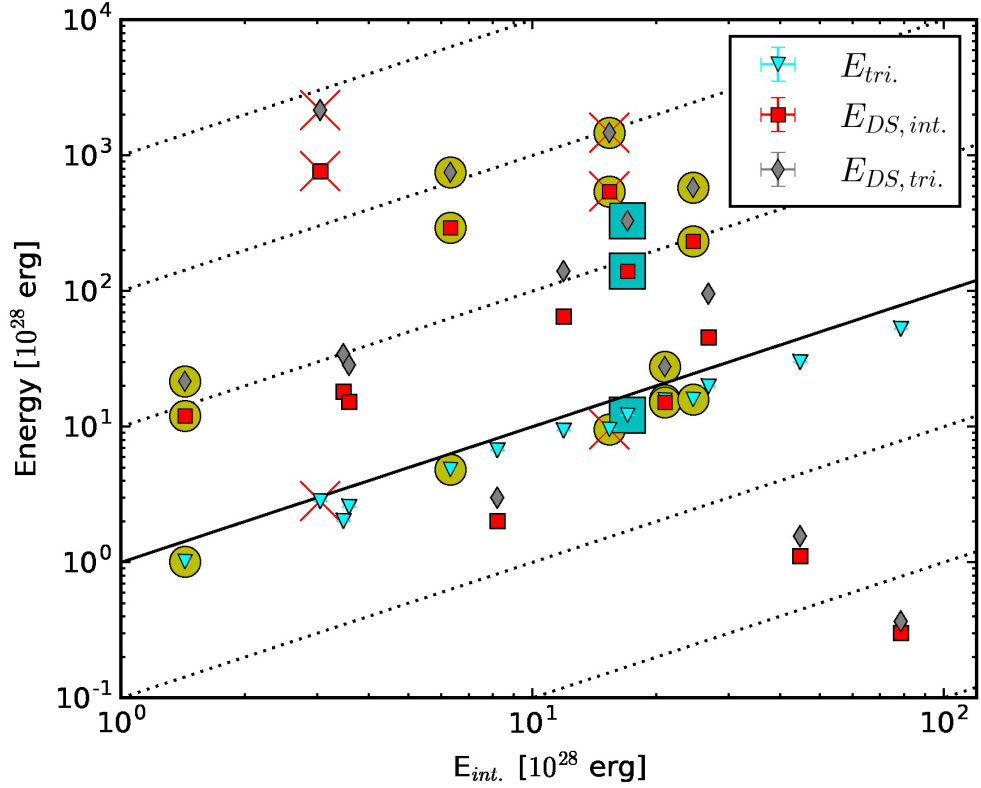} 
\caption{
Comparison of the flare energies listed in Table 4. 
Events listed as `poor' in the SOHO/LASCO catalog are marked by a yellow circle around the data point.
Events that followed a non-exponential shape are marked with a red X.  
The blue square marks the event that had multiple signals, but the most exponential shaped event was chosen.
The black line traces a 1:1 correspondence while the dotted lines represent order of magnitude differences from the solid line.  }
\label{fig:energies}
\end{figure}

Table 4 lists a summary of the parameters obtained by the GOES analysis and the energy values obtained by equation (\ref{eq:Combined}).
Figure \ref{fig:energies} shows the relationship between the integrated X-ray flare energy $E_{int.}$ and the other various types of energies when plotted on a log-log scale. 
The black line traces a 1-to-1 correspondence.  
The energy defined by using a triangular flare decay shape $E_{tri.}$ traces just under the 1-to-1 line highlighting the similarities of the two energies.   
The other trend is the closeness of the paired methods.
$E_{int.}$ is close to $E_{tri.}$ and E$_{DS,int.}$ is very similar to E$_{DS,tri.}$.

\begin{deluxetable}{lc | ccccc | cc | cc}
	\rotate{}
	\tabletypesize{\footnotesize}
	\tablecaption{Parameters Determined from GOES Analysis}
	\tablewidth{0pt}
	\tablehead{	\vspace{ -1.21cm}  }
	\startdata 
	GOES & Date & Duration & Peak Flux & Fluence & Temp. & Scale Height & E$_{int.}$  & E$_{tri.}$ &  E$_{DS,int.}$ & E$_{DS,tri.}$   \\
	Class & of Event & [s] & [$\mu$W/m$^{2}$] & [mJ/m$^{2}$] & [10$^{6}$ K] & [10$^{10}$ cm] & [10$^{28}$ erg] & [10$^{28}$ erg] & [10$^{28}$ erg] & [10$^{28}$ erg]  \\  \cline{1-11}
	X\tablenotemark{\dagger'} & 2012-07-06 		& 540 & 161  & 60 & 5.02 $\pm$ 0.56 & 2.5 $\pm$ 0.28 & 17 & 12 & 136 $\pm$ 603 & 318 $\pm$ 2457 \\
	X & 2013-11-08 						& 420 & 160 & 42 &  2.84 $\pm$ 0.11 & 1.42 $\pm$ 0.06 & 12 & 9 & 63 $\pm$ 280 & 137 $\pm$ 1042\\
	X & 2014-04-25 						& 1080 & 198 & 159 & 3.31 $\pm$ 0.1 & 1.66 $\pm$ 0.05 & 45 & 30 & 1.1 $\pm$ 4.6 & 1.5 $\pm$ 11   \\  \cline{1-11}
	M & 2011-01-28 						& 720 & 20 & 12 & 3.07 $\pm$ 0.07 & 1.54 $\pm$ 0.03 & 3.5 & 2.0 & 18 $\pm$ 77 & 33 $\pm$ 249   \\ 
	M\tablenotemark{*}\tablenotemark{\dagger}  & 2011-02-13 & 720 & 95 & 55 & 3.86 $\pm$ 0.27 & 1.94 $\pm$ 0.14 & 15 & 10 & 528 $\pm$ 2368 & 1431 $\pm$ 11267  \\  
	M\tablenotemark{*} & 2012-03-17 			& 360 & 20 & 5 & 4.82 $\pm$ 0.37 & 2.42 $\pm$ 0.19 & 1.4 & 1 & 12 $\pm$ 51 & 21 $\pm$ 155  \\
	M\tablenotemark{\dagger} & 2012-06-03		& 420 & 48 & 11 & 3.59 $\pm$ 0.22 & 1.8 $\pm$ 0.11 & 3.0 & 2.8 & 747 $\pm$ 3358 & 2101 $\pm$ 16630 \\
	M & 2013-05-02 						& 1140 & 16 & 13 & 2.84 $\pm$ 0.15 & 1.42 $\pm$ 0.08 & 3.6 & 2.6 & 15 $\pm$ 65 & 28 $\pm$ 205  \\ 
	M\tablenotemark{*} & 2013-10-24 			& 840 & 134 & 75 & 9.58 $\pm$ 1.43 & 4.8 $\pm$ 0.72 & 21 & 16 & 15 $\pm$ 64 & 27 $\pm$ 201   \\ 
	M\tablenotemark{*} & 2014-01-08 			& 660 & 52 & 22 & 3.7 $\pm$ 0.56 & 1.85 $\pm$ 0.28 & 6.3 & 4.9 & 295 $\pm$ 1322 & 752 $\pm$ 5876  \\
	M\tablenotemark{*} & 2014-02-20 			& 2580 & 44 & 87 & 4.2 $\pm$ 0.34 & 2.1 $\pm$ 0.17 & 25 & 16 & 227 $\pm$ 1011 & 562 $\pm$ 4368  \\
	M & 2014-03-20 						& 1920 & 25 & 29 & 3.12 $\pm$ 0.14 & 1.56 $\pm$ 0.07 & 8.2 & 6.7 & 2 $\pm$ 8 & 3 $\pm$ 21 \\
	M & 2014-11-03 						& 1500 & 94 & 95 & 2.72 $\pm$ 0.05 & 1.36 $\pm$ 0.03 & 27 & 20 & 44 $\pm$ 195 &  92 $\pm$ 699 \\
	M & 2014-12-17  						& 3000 & 125 & 28 & 3.59 $\pm$ 0.12 & 1.8 $\pm$ 0.06 & 79 & 53 & 0.3 $\pm$ 1.3 & 0.4 $\pm$ 2.5  \\
	\enddata
	\tablecomments{ 
	Energies are calculated using the average temperature T = $4.0 \pm 0.5 \times 10^{6}$ K  and average scale height $H_{0} = 2.0 \pm 0.2 \times 10^{10}$ cm.
	Flare Duration is defined as the time difference between initial brightening and the point where the flare has decayed to half of the maximum flux.  
	}
\end{deluxetable}

Due to the similarities in the parameters A and $\gamma$ used in equation (\ref{eq:Mcme}) by \citet{Aarnio2012} and \citet{Drake2013}, a similar value for mass is expected given the same input energy.
Section \ref{sec:mass} discusses this more thoroughly.

E$_{DS,tri.}$ and E$_{DS,int.}$, described by equation (\ref{eq:Combined}), describes the predicted flare energy given a velocity measurement from a type II burst.
The predicted values tend to agree within a rough order of magnitude with E$_{int.}$. 
However, there are large errors associated with them which make them unreliable.  

The derivation of equation (\ref{eq:Combined}) uses the assumption that the masses determined by (\ref{eq:Mcme}) and (\ref{eq:equipar}) provide the same value.
In general, these masses will have large errors and will only approximately agree with each other; this will be discussed in more detail in Section \ref{sec:mass}.
Therefore the large errors in mass and their discrepancies create the large errors found for E$_{DS,tri.}$ and E$_{DS,int.}$.  
This puts into question the applicability of equation (\ref{eq:Combined}) until more rigorous constraints are achieved than what is currently available.
 
\subsection{CME Velocity from Dynamic Spectra}
 
\begin{figure}
\centering
\includegraphics[angle=-90,scale=0.5,clip=true, trim=0cm 0cm 0cm 0cm]{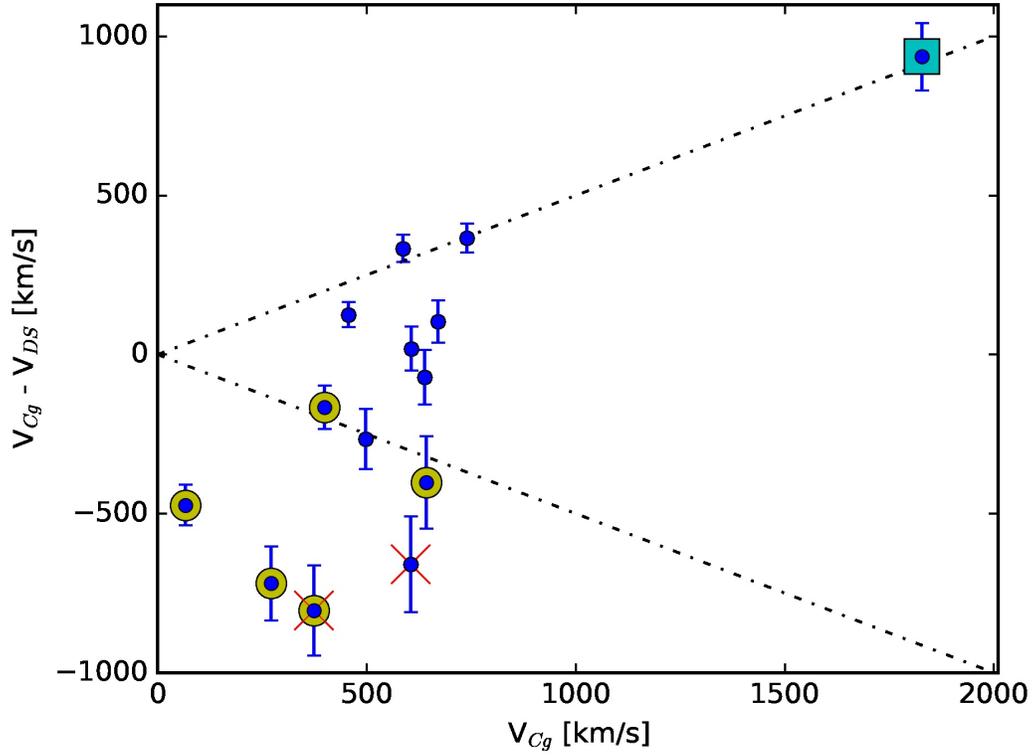} 
\caption{
Difference between velocities measured from coronagraph observations and inferred from dynamic spectra analysis, as a function of $V_{Cg}$.
The dashed lines represent $\pm$50\% of $V_{Cg}$. 
Values come from Table 5.
Events listed as `poor' in the SOHO/LASCO catalog are marked by a yellow circle around the data point.
Events that followed a non-exponential shape are marked with a red X.  
The blue square marks the event that had multiple signals, but the most exponential shaped event was chosen.
}
\label{fig:vels}
\end{figure}

Table 5 shows the evaluated velocities for the average temperature case.  
The velocities of the LASCO/SOHO ($V_{Cg}$) and dynamic spectra ($V_{DS}$) tend to disagree beyond the measurement errors.
Figure \ref{fig:vels} displays the difference between $V_{Cg}$ and $V_{DS}$ as a function of $V_{Cg}$.

Events listed as `poor' in the SOHO/LASCO catalog are marked by a yellow circle around the data point.
Events that followed a non-exponential shape are marked with a red X.  
The blue square marks the event that had multiple signals, but the most exponential shaped event was chosen.
The most egregious points are typically marked as being a poor event or follow a non-exponential shape.
This is to be expected.  
Any points that are non-exponential tend to be more steeply shaped than would describe an exponential.  
This is due to the local density not following a barometric model.
\citet{Leblanc1999} and \citet{Guhathakurta1999} demonstrate that the solar corona varies more steeply then would be expected by a barometric corona.  
Since events are modeled as moving though a barometric atmosphere, the modeled shock velocity is large to account for lower assumed density gradient. 
Additionally, any poor events would be expected to have poorer results when trying to measure its properties. 

Good events, that are described well by an exponential model have an average absolute discrepancy of 276 $\pm$ 293 km/s.
An underestimation is anticipated because the measurement is of the type II bursts shock speed rather than the CME directly.  
The shock surface is constrained to a maximum velocity of the CME, but the shock surface can move relative to the leading edge of the CME which would lower its measured velocity.  

\begin{figure}
\centering
\includegraphics[angle=-90,scale=0.5,angle=0,clip=true, trim=0cm 0cm 0cm 0cm]{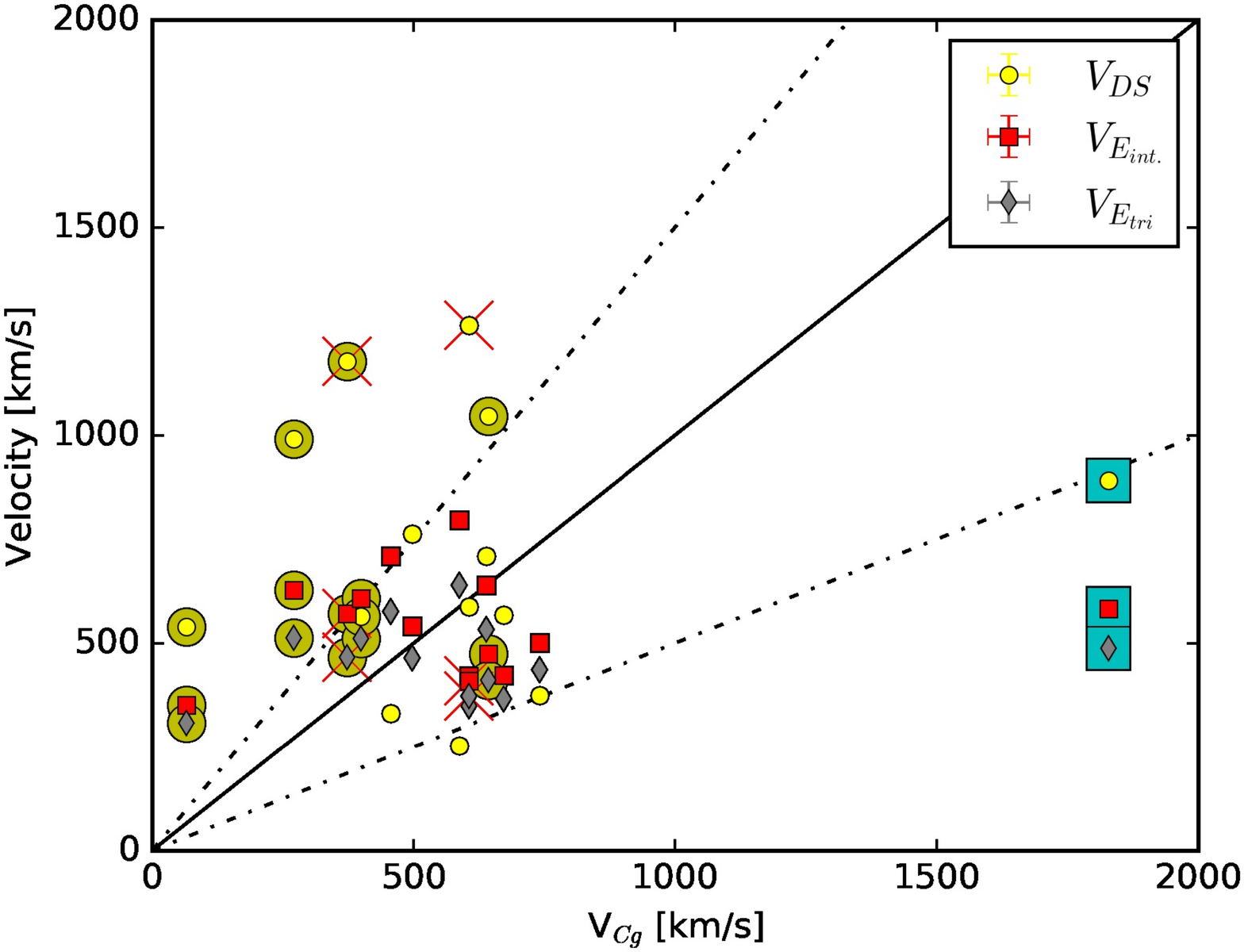} 
\caption{
Velocities derived from analysis against the velocity determined by coronagraphic analysis. 
Table 5 lists specific values.
Error bars are omitted from this plot.  
Events listed as `poor' in the SOHO/LASCO catalog are marked by a yellow circle around the data point.
Events that followed a non-exponential shape are marked with a red X.  
The blue square marks the event that had multiple signals, but the most exponential shaped event was chosen.
The black line traces a 1:1 correspondence while the dashed lines represent $\pm$50\% from the solid line. }
\label{fig:velocities}
\end{figure}

The CME velocities predicted by flare observations ($V_{E_{int.}}$ and $V_{E_{tri.}}$) have large errors associated to them and roughly agree with observation.   
$V_{E_{tri.}}$ always predicts a slightly lower velocity, due to its slightly lower input energy, than $V_{E_{int.}}$.  
While the exact predicted value is likely not to be trusted, its value is a reasonably measure for an order of magnitude estimate of velocities. 
Figure \ref{fig:velocities} presents the derived velocities as compared to the observed ($V_{Cg}$) velocity.

\subsection{Determination of CME Mass}
\label{sec:mass}

\begin{figure}
\centering
\includegraphics[scale=.6,clip=true, trim=0cm 0cm 0cm 0cm]{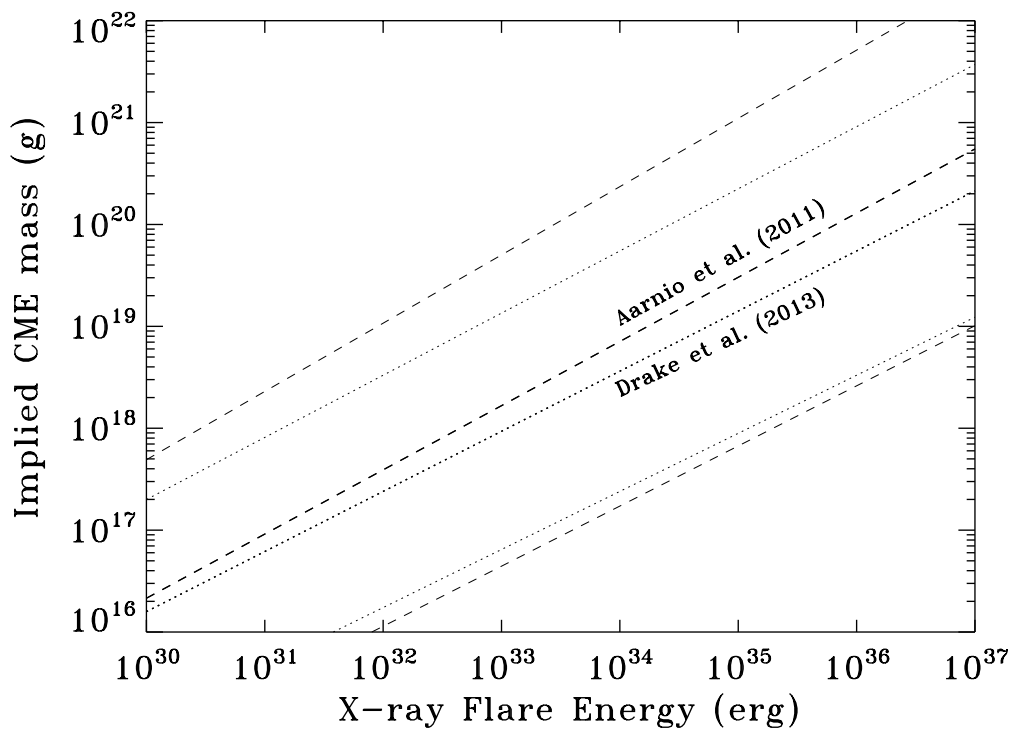} 
\caption{
Plot of Aarnio and Drake empirical relationship compared to each other and an indication of the uncertainty in the equations themselves.
CME mass determined from the Aarnio relation and uncertainty range are shown with dashed lined  while masses determined using the Drake relation are shown using dotted lines.
 }
\label{fig:AD_plot}
\end{figure}

The masses derived from the empirical relationships ($M_{int.}$ and $M_{tri}$) are plotted in Figure \ref{fig:AD_plot}.
The ability to derive a mass from a flare energy using these relationships comes with a large uncertainty.   
This large uncertainly (larger than 100\%) is prohibitive when determining a lower bound for the CME mass.
The application of these equations is likely better suited to determine the average mass for a given event energy rather than its accuracy in a case-by-case evaluation. 

As mentioned above, the X-ray energy is defined differently for the Aarnio and Drake equations.  
This manifests as a higher flare energy parameter for the Aarnio relation. 
As Figure \ref{fig:AD_plot} demonstrates, since both the Aarnio and Drake equations are so similar, this implies that the mass derived from Aarnio ($M_{tri.}$) will be larger than Drake ($M_{int.}$) for a particular event.

\begin{figure}
\centering
\includegraphics[angle=-90,scale=0.5,angle=0,clip=true, trim=0cm 0cm 0cm 0cm]{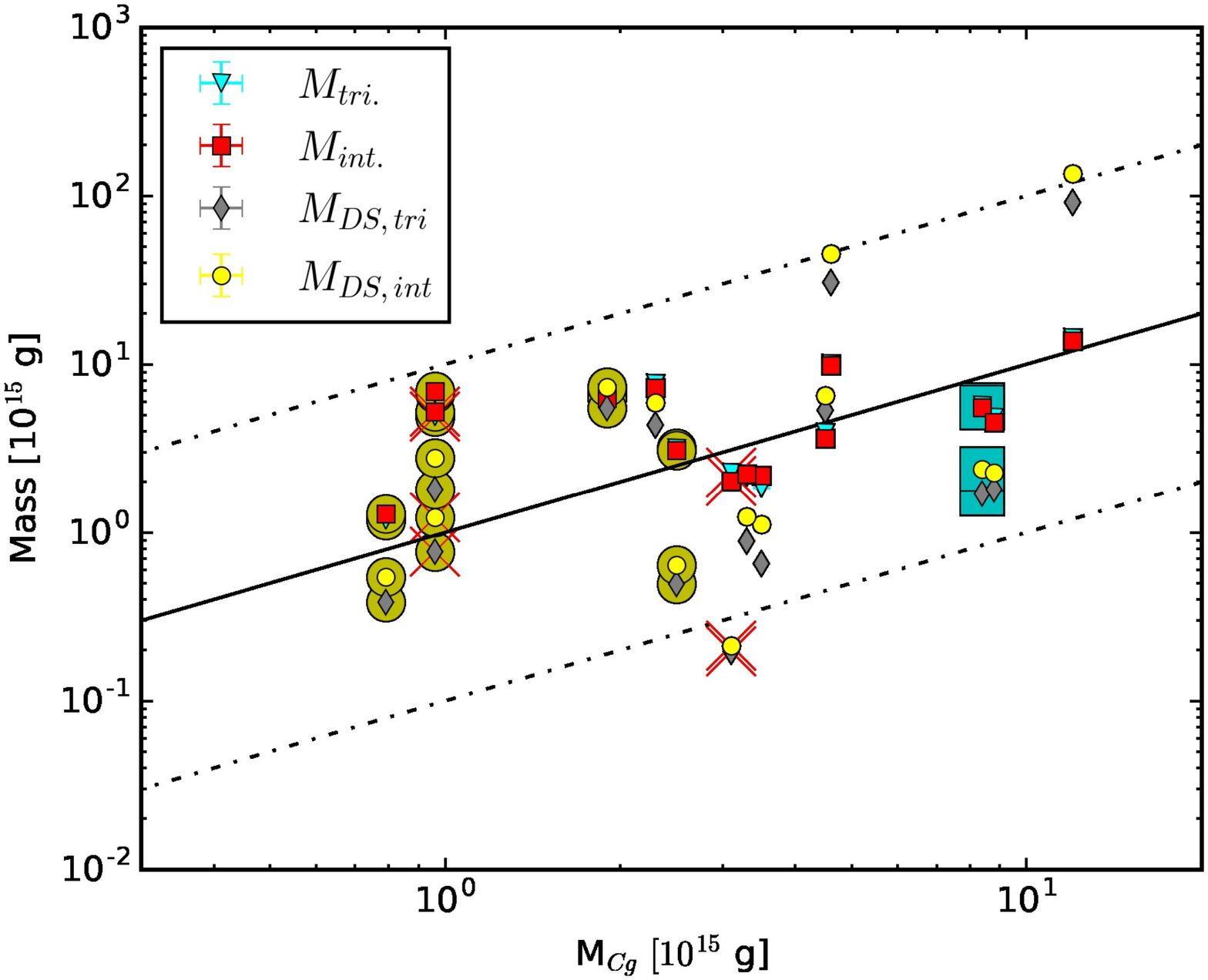} 
\caption{
Masses derived from analysis against the mass determined by coronagraphic analysis. 
Table 5 lists specific values.
Error bars are omitted from this plot.  
Events listed as `poor' in the SOHO/LASCO catalog are marked by a yellow circle around the data point.
Events that followed a non-exponential shape are marked with a red X.  
The blue square marks the event that had multiple signals, but the most exponential shaped event was chosen.
The black line traces a 1:1 correspondence while the dotted lines represent order of magnitude differences from the solid line. }
\label{fig:masses}
\end{figure}

Table 5 shows the evaluated masses for the unique and average temperature situations respectively.  
Figure \ref{fig:masses} displays this information graphically. 
Comparing the best value predictions to the coronagraphic measurement ($M_{Cg}$), both $M_{int.}$ and $M_{tri.}$ perform equally well. 
In fact, all masses are held within an order of magnitude of $M_{Cg}$.  
The large errors still put into question the reliance of these values in observations.  

The discrepancy in the masses ($M_{DS,tri.}$ and $M_{DS,int.}$) derived by equation (\ref{eq:equipar}) is primarily due to the small difference between $E_{int.}$ and $E_{tri.}$..
Since these masses are simply a scaled and converted Energy, the disparity in mass directly correlates to the disparity in the energy used for calculations. 
$M_{DS,int.}$ very often is nearly two orders of magnitude below $M_{DS,tri.}$ and $M_{Cg}$.  
This suggests that the product of the factors $f_{GOES}$ and $\epsilon$ is not yet refined enough to be able to be used in calculations.  

\begin{figure}
\centering
\includegraphics[angle=-90,scale=0.5,angle=0,clip=true, trim=0cm 0cm 0cm 0cm]{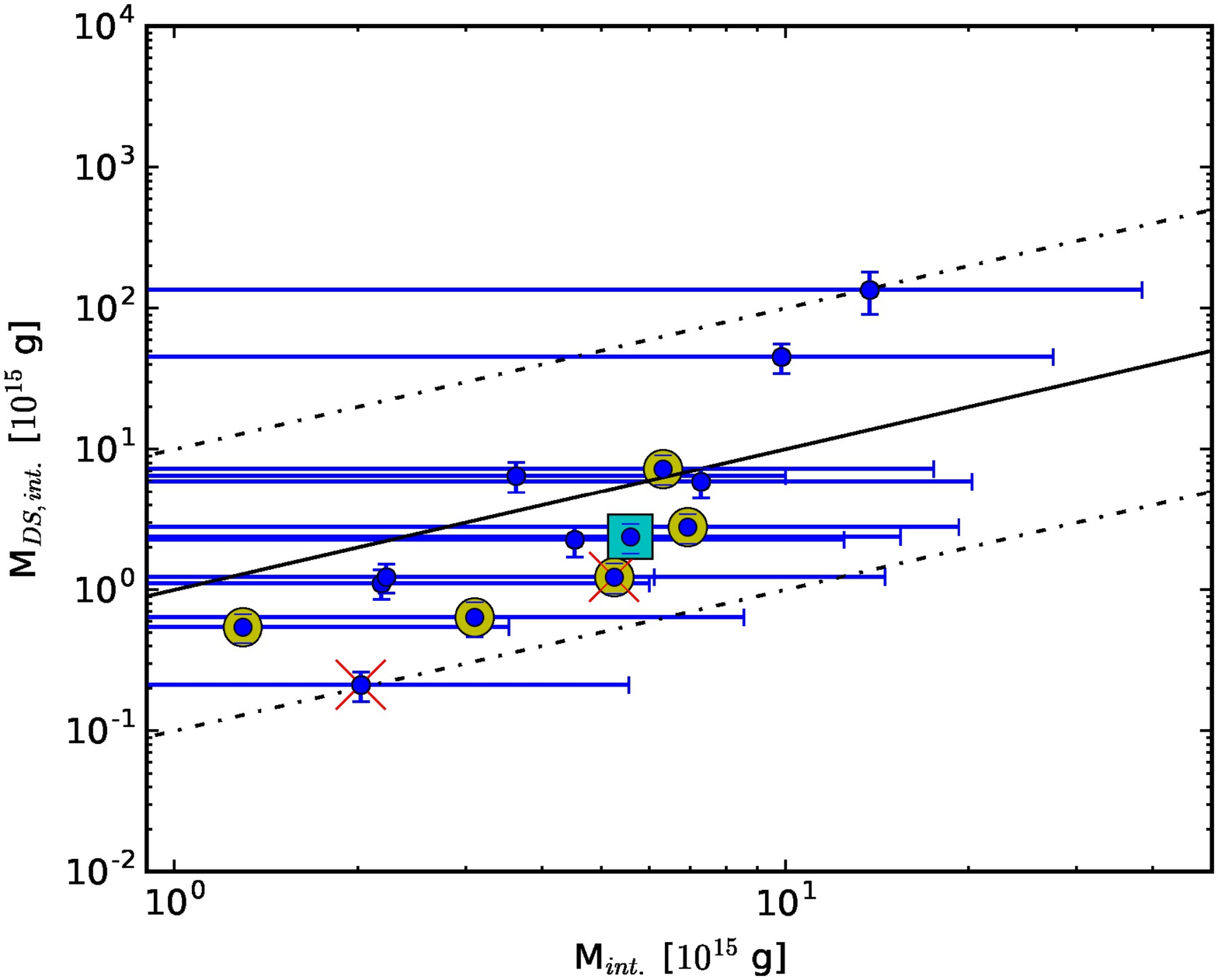} 
\caption{
Plots the mass determined by Drakes equation $M_{int.}$ vs the mass determined from the equiparition relationship $M_{DS,int.}$.
The agreement of these two values (the closer they are to the black line) determines if they can be assumed to be equal for use equations (\ref{eq:Combined}) and (\ref{eq:VCombined}).
Events listed as `poor' in the SOHO/LASCO catalog are marked by a yellow circle around the data point.
Events that followed a non-exponential shape are marked with a red X.  
The blue square marks the event that had multiple signals, but the most exponential shaped event was chosen.
The black line traces a 1:1 correspondence while the dotted lines represent order of magnitude differences from the solid line.
 }
\label{fig:mvm}
\end{figure}

The agreement between $M_{int.}$ and $M_{DS,int.}$ is important for the determination of $E_{DS,int.}$ and is plotted in Figure \ref{fig:mvm}.  
As mentioned above, $E_{DS,int.}$ is derived by the assumption that these masses are equivalent.
Figure \ref{fig:mvm} shows how $M_{DS,int.}$ tends to be approximately the same but, $M_{DS,int.}$ does not hold a constant relationship to $M_{int.}$
Equation (\ref{eq:Combined}) would need an additional variable scaling factor to allow for the equivalency.  
As $E_{DS,int.}$ already contains large errors, this additional inconsistency truly put this value into question.  
 
\begin{deluxetable}{lc | cccc | ccccc}
	\rotate{}
	\tabletypesize{\footnotesize}
	\tablecaption{Velocity and Mass}
	\tablewidth{0pt}
	\tablehead{\vspace{ -1.21cm} }
	\startdata 
	GOES & Data & V$_{Cg}$  & V$_{DS}$  & V$_{E_{int.}}$  & V$_{E_{tri.}}$  & M$_{Cg}$ & M$_{tri.}$ & M$_{int.}$ & M$_{DS,tri.}$  & M$_{DS,int.}$ \\
	Class & of Event &  [km/s] &  [km/s] &  [km/s] &  [km/s] & [10$^{15}$ g] & [10$^{15}$ g] & [10$^{15}$ g] & [10$^{15}$ g] &   [10$^{15}$ g] \\ \cline{1-11}
	X\tablenotemark{\dagger'} & 2012-07-06 		& 1828 	& 897 $\pm$ 114 	& 582 $\pm$ 516 & 488 $\pm$ 791	& 8.4 & 5.7 $\pm$ 15.5 	& 5.6 $\pm$ 9.9  	& 1.7 $\pm$ 0.4 	& 2.3 $\pm$ 0.56 \\ 
	X & 2013-11-08 						& 497 	& 767 $\pm$ 101	& 541 $\pm$ 478 & 465 $\pm$ 732	& 8.8 & 4.8 $\pm$ 13.1 	& 4.5 $\pm$ 8	 	& 1.8 $\pm$ 0.4 	& 2.4 $\pm$ 0.6 \\ 
	X & 2014-04-25 						& 456  	& 333 $\pm$ 42 	& 710 $\pm$ 634 & 576 $\pm$ 977	& 4.6 & 10 $\pm$ 28 	& 9.9 $\pm$ 17.6 	& 30 $\pm$ 7	 	& 45 $\pm$ 11  \\  \cline{1-11}
	M & 2011-01-28 						& 606 	& 591 $\pm$ 75	& 420 $\pm$ 367 & 350 $\pm$ 556	& 3.5 &  1.8 $\pm$ 4.9 	& 2.2 $\pm$ 3.8 	& 0.65 $\pm$ 0.15 	& 1.1 $\pm$ 0.3  \\
	M\tablenotemark{*}\tablenotemark{\dagger} & 2011-02-13 & 373 & 1185 $\pm$ 153 & 570 $\pm$ 505 & 466 $\pm$ 772	& 0.96 & 4.9 $\pm$ 13 & 5.3 $\pm$ 9.3 &  0.8 $\pm$ 0.2 	& 1.2 $\pm$ 0.3\\ 
	M\tablenotemark{*} & 2012-03-17 			& 66 		& 542 $\pm$ 68 	& 350 $\pm$ 304 & 308 $\pm$ 459	& 0.79 & 1.2 $\pm$ 3.1 	& 1.3 $\pm$ 2.2  	& 0.39 $\pm$ 0.1 	& 0.55 $\pm$ 0.13 \\ 
	M\tablenotemark{\dagger} & 2012-06-03 		& 605 	& 1272 $\pm$ 161	& 409 $\pm$ 357 & 372 $\pm$ 544	& 3.1 & 2.3 $\pm$ 6 		& 2 $\pm$ 3.5		& 0.2 $\pm$ 0.05 	& 0.2 $\pm$ 0.05 \\
	M & 2013-05-02 						& 671 	& 570 $\pm$ 72	& 423 $\pm$ 370 & 366 $\pm$ 562	& 3.3 & 2.14 $\pm$ 5.7	& 2.2 $\pm$ 3.9  	& 0.9 $\pm$ 0.21	& 1.24 $\pm$ 0.3 \\
	M\tablenotemark{*} & 2013-10-24 			& 399 	& 568 $\pm$ 73 	& 608 $\pm$ 539 & 512 $\pm$ 829	& 1.9 & 6.7 $\pm$ 18.3 	& 6.3 $\pm$ 11.2 	& 5.5 $\pm$ 1.3	& 7.3 $\pm$ 1.7 \\
	M\tablenotemark{*} & 2014-01-08 			& 643 	& 1045 $\pm$ 145	& 475 $\pm$ 417 & 411 $\pm$ 637	& 2.5 & 3.2 $\pm$ 8.6 	& 3.1 $\pm$ 5.4 	& 0.5 $\pm$ 0.1 	& 0.6 $\pm$ 1.8 \\ 
	M\tablenotemark{*} & 2014-02-20 			& 271 	& 997 $\pm$ 126 	& 628 $\pm$ 558 & 513 $\pm$ 856	& 0.96 & 6.8 $\pm$ 18.4 	& 6.9 $\pm$ 12.3 	& 1.8 $\pm$ 0.4 	& 2.8 $\pm$ 0.7 \\ 
	M & 2014-03-20 						& 740 	& 376 $\pm$ 47 	& 502 $\pm$ 441 & 437 $\pm$ 675	& 4.5 & 3.9 $\pm$ 10.5 	& 3.6 $\pm$ 6.4  	& 5.3 $\pm$ 1.3 	& 6.5 $\pm$ 1.6 \\ 
	M & 2014-11-03 						& 638 	& 713 $\pm$ 92 	& 639 $\pm$ 568 & 533 $\pm$ 874	& 2.3 & 7.7 $\pm$ 21 	& 7.3 $\pm$ 13 	& 4.3 $\pm$ 1.1	& 5.9 $\pm$ 1.4 \\
	M & 2014-12-17 						& 587 	& 255 $\pm$ 44 	& 797 $\pm$ 715 & 639 $\pm$ 1105	& 12 & 14.3 $\pm$ 40 	& 13.7 $\pm$ 25	& 91 $\pm$ 30 		& 136 $\pm$ 45 \\ 
	\enddata
	\tablecomments{
	Average Temperature T = $4.0 \pm 0.5 \times 10^{6}$ K  and average scale height $H_{0} = 2.0 \pm 0.2 \times 10^{10}$ cm.
	Errors are low for M$_{DS,tri.}$ and M$_{DS,int.}$ because the error's of $\epsilon$ and $f_{GOES}$ are assumed to be 0.  
	}
\end{deluxetable}
 
\subsection{Determination of Kinetic Energy}

\begin{figure}
\centering
\includegraphics[angle=-90,scale=0.5,angle=0,clip=true, trim=0cm 0cm 0cm 0cm]{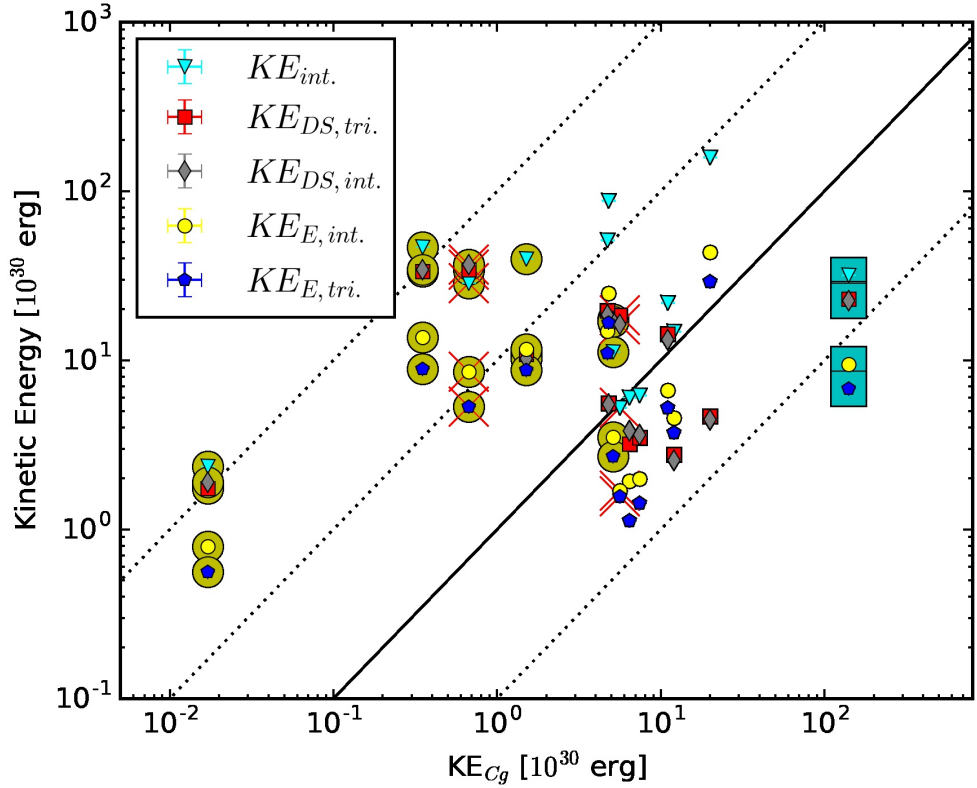} 
\caption{
Kinetic energies derived from analysis against the kinetic energy determined by coronagraphic analysis. 
Table 7 lists specific values.
Error bars are omitted from this plot.  
Events listed as `poor' in the SOHO/LASCO catalog are marked by a yellow circle around the data point.
Events that followed a non-exponential shape are marked with a red X.  
The blue square marks the event that had multiple signals, but the most exponential shaped event was chosen.
The black line traces a 1:1 correspondence while the dotted lines represent order of magnitude differences from the solid line.
 }
\label{fig:KEs}
\end{figure}

The kinetic energies derived are listed in Table 6.
These values are presented graphically in Figure \ref{fig:KEs}.
The errors are larger when considering the average temperature scenario, however the errors are already quite large such that the additional error contribution does not change the interpretation significantly.  

Errors for all Kinetic Energies (with the exception of KE$_{E,tri.}$ and KE$_{E,int.}$ which have low errors,$\delta_{\epsilon} = \delta_{f} = 0$) are all larger than 100\%.
The best values also tend to have  variation between them but are typically within an order of magnitude. 
Kinetic energy evaluation may still best utilized as an order or magnitude or upper bound limiting tool rather than an accurate assessment of an event's kinetic energy until mass and/or velocity measurements can be more tightly constrained.  

\begin{figure}
\centering
\includegraphics[angle=-90,scale=0.5,angle=0,clip=true, trim=0cm 0cm 0cm 0cm]{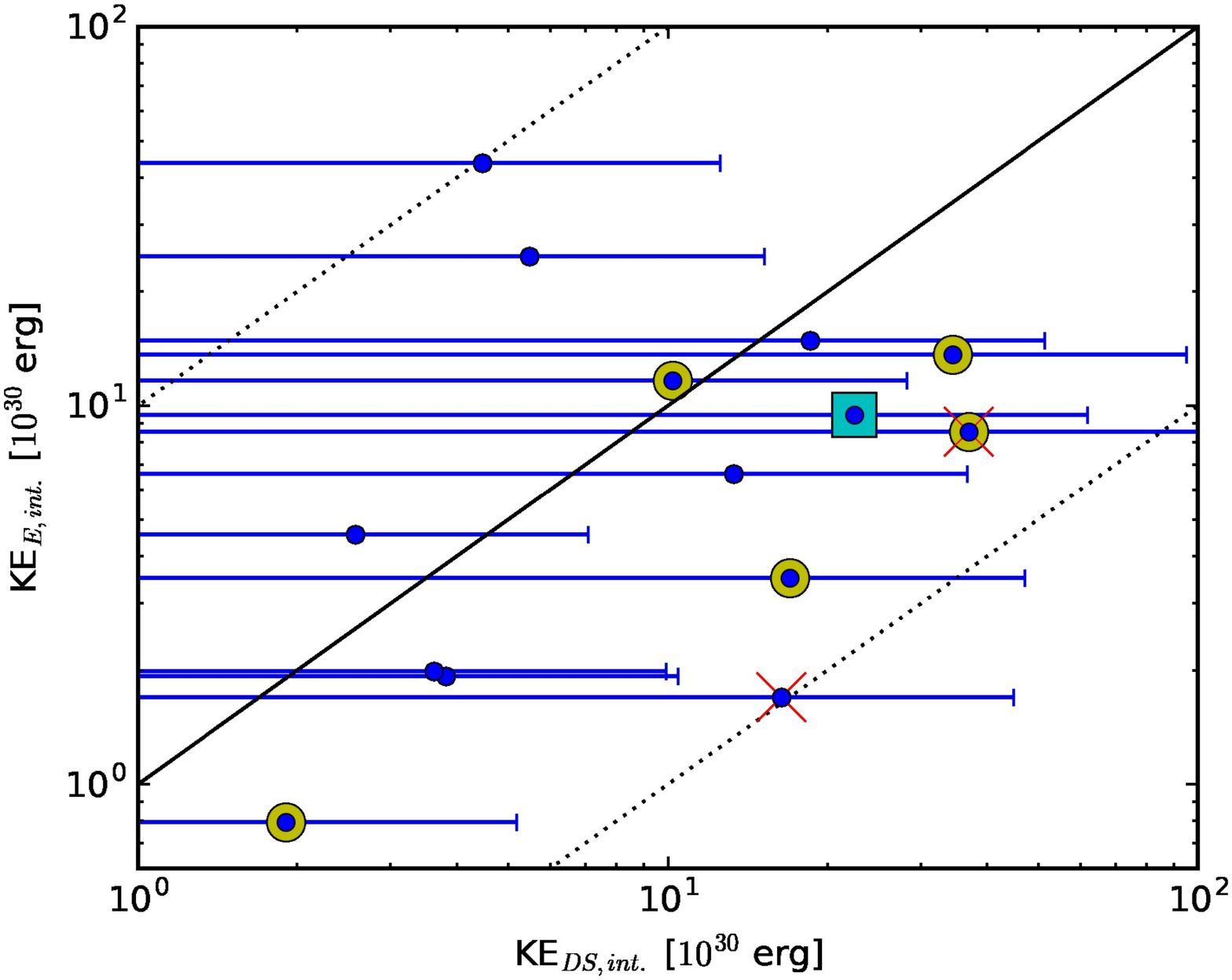} 
\caption{
Both sides of the equipartition relationship equation.  
The closer to the black line, the more valid the equation and thus it is more reliable for later calculations.  
Errors are present in Y-axis, they are just minimal as the errors of  $\epsilon$ and $f_{GOES}$ are assumed to be zero ($\delta_{\epsilon} = \delta_{f} = 0$).
Table 6 lists specific values.
Events listed as `poor' in the SOHO/LASCO catalog are marked by a yellow circle around the data point.
Events that followed a non-exponential shape are marked with a red X.  
The blue square marks the event that had multiple signals, but the most exponential shaped event was chosen.
The black line traces a 1:1 correspondence while the dotted lines represent order of magnitude differences from the solid line. 
}
\label{fig:kvk}
\end{figure}

$KE_{DS,int.}$ and $KE_{E,int.}$ represent both sides of the equipartition relationship and is plotted in Figure \ref{fig:kvk}.
This is an additional test to verify that the mass ($M_{DS,int}$) derived from it is indeed accurate.  
There is a larger spread in the correlation than was found in the associated mass relationship.  
The large errors allow that these two agree within error but does not support the equivalence required for the two values.

\begin{deluxetable}{lc | ccccccc}
	\tabletypesize{\footnotesize}
	\tablecaption{Kinetic Energy [10$^{30}$ erg]}
	\tablewidth{0pt}
	\tablehead{	\vspace{ -1.21cm}  }
	\startdata 
	GOES Class & Date of Event & KE$_{Cg}$ & KE$_{int.}$ & KE$_{DS,tri.}$ & KE$_{DS,int.}$ & KE$_{E,tri.}$ & KE$_{E,int.}$ \\
	\cline{1-8}
	X\tablenotemark{\dagger'} & 2012-07-06 			& 140 	& 32 $\pm$ 89  		& 23 $\pm$ 62  		& 22 $\pm$ 40	 	& 6.8 $\pm$ 0.03  		& 9.4 $\pm$ 0.0007     \\
	X & 2013-11-08 							& 11 		& 22 $\pm$ 61 		   	& 14 $\pm$ 38	         	& 13	$\pm$ 23	 	& 5.2 $\pm$ 0.02 		& 6.6 $\pm$ 0.0007        \\
	X & 2014-04-25 							& 4.8	 	& 88 $\pm$ 248 		& 6 $\pm$ 15     		& 5.4 $\pm$ 9.7 	& 16.7 $\pm$ 0.07	 	& 24.8 $\pm$ 0.0008  \\                     \cline{1-8}
	M & 2011-01-28 							& 6.4 	& 6 $\pm$ 17 			& 3.2 $\pm$ 8.4		& 3.8 $\pm$ 6.6 	& 1.1 $\pm$ 0.01 		& 1.9 $\pm$ 0.0002 \\
	M\tablenotemark{*}\tablenotemark{\dagger}  & 2011-02-13 & 0.67 & 29 $\pm$ 80 		& 34 $\pm$ 92   		& 36 $\pm$ 65		& 5.3 $\pm$ 0.03 		& 8.5 $\pm$ 0.0006 \\ 
	M\tablenotemark{*} & 2012-03-17 				& 0.017 	& 2.4 $\pm$ 6.5   		& 1.7 $\pm$ 4.5 		& 1.9 $\pm$ 3.3 	& 0.6 $\pm$ 0.01  		& 0.79 $\pm$ 0.00002     \\
	M\tablenotemark{\dagger} & 2012-06-03 			& 5.6 	& 5 $\pm$ 15 			& 18 $\pm$ 48  		& 16 $\pm$ 28	  	& 1.6 $\pm$ 0.01 		& 1.7 $\pm$  0.0004   \\
	M & 2013-05-02 							& 7.4 	& 6 $\pm$ 17 			& 3 $\pm$ 9 		   	& 3.6 $\pm$ 6.3 	& 1.4 $\pm$ 0.02 		& 2 $\pm$ 0.0002 \\
	M\tablenotemark{*} & 2013-10-24 				& 1.5 	& 40 $\pm$ 112         	& 11 $\pm$ 29   		& 10 $\pm$ 18	 	& 8.8 $\pm$ 0.04 		& 11.7 $\pm$ 0.0007\\
	M\tablenotemark{*} & 2014-01-08 				& 5.1 	& 11 $\pm$ 31 			& 17 $\pm$ 47 			& 17 $\pm$ 30	 	& 2.7 $\pm$ 0.02 		& 3.5 $\pm$ 0.0004    \\	
	M\tablenotemark{*} & 2014-02-20 				& 0.35 	& 47 $\pm$ 132 		& 33 $\pm$ 91 			& 34 $\pm$ 61	 	& 8.9 $\pm$ 0.08 		& 13.7 $\pm$ 0.0004    \\
	M & 2014-03-20 							& 12 		& 15 $\pm$ 41 		  	& 3 $\pm$ 7 			& 2.5 $\pm$ 4.5 	& 3.7 $\pm$ 0.04 		& 4.6 $\pm$  0.0002  \\
	M & 2014-11-03 							& 4.7 	& 51 $\pm$ 144 		& 19 $\pm$ 53	 		& 18 $\pm$ 33	 	& 11 $\pm$  0.07 		& 14.9 $\pm$ 0.0006   \\
	M & 2014-12-17 							& 20 		& 158 $\pm$ 451 		& 5 $\pm$ 13			& 4 $\pm$ 8	 	& 29.3 $\pm$ 0.15 		& 43.6 $\pm$ 0.0007    \\
	\enddata
	\tablecomments{
	Average Temperature T = $4.0 \pm 0.5 \times 10^{6}$ K  and average scale height $H_{0} = 2.0 \pm 0.2 \times 10^{10}$ cm.
	Errors are low for KE$_{E,tri.}$ and KE$_{E,int.}$ because the error's of $\epsilon$ and $f_{GOES}$ are assumed to be 0.  
	}
\end{deluxetable}

\section{Conclusion}

The comparison between white light coronagraphic data to a combination of GOES X-ray and BIRS radio data provides a way to test methodology of interpreting CME observations on stellar objects.
Velocity measurements, derived from radio dynamic spectra, provide reasonably accurate results, particularly when pre-flare temperatures are accessible.  
Good events, that are described well by an exponential model have an average discrepancy of 276 $\pm$ 293 km/s.

Mass determined through empirical or equipartion relationships tend to agree within error, but this is due to the large errors.
This is useful in the stellar situation when trying to achieve initial constraints regarding CME mass but the empirical equations are still not yet strong enough to determine better than order of magnitude values. 

$E_{DS,int.}$ is derived by the assumption that these masses are equivalent and the agreement between $M_{int.}$ and $M_{DS,int.}$ is important for this relation to hold true. 
This poor equivalence suggests that the empirical relationships used to determine mass are not refined enough for calculations.
It also suggests that the product of the factors $f_{GOES}$ and $\epsilon$ is not yet refined enough to be able to be used in calculations.

The best values of kinetic energy tend to be close but large errors deriving allow for all values to agree with each other within error.  
The applicability of the equipartition equation (\ref{eq:equipar}) to solve for $M_{DS,int}$ still poses some issues. 
However, kinetic energy evaluation may still best utilized as an order or magnitude or upper bound limiting tool than an accurate assessment of an events kinetic energy until mass and/or velocity measurements can be more tightly constrained.

Overall, this methodology provides reasonable velocity constraints while also providing order of magnitude constraints for mass and kinetic energies.  
Translating this to the stellar case should come with little difficulty and should provide a reliable and valuable means to constrain observed events where no other constraints exist.  

\acknowledgements
MKC and RAO acknowledges funding support from NSF AST-1412525 for the project on which this paper is based.  
This CME catalog is generated and maintained at the CDAW Data Center by NASA and The Catholic University of America in cooperation with the Naval Research Laboratory. SOHO is a project of international cooperation between ESA and NASA. 

\appendix
\section{Error Analysis}
\label{sec:EA}

The error for the GOES x-ray light curve measurements is assumed to follow a Poisson distribution.   
The flux (F) is $F = \frac{counts}{t_{exp}} \lambda$ with error: $\sigma_{F} = \frac{\sqrt{counts}}{t_{exp}} \lambda$.  
The exposure time is $t_{exp}$ and $\lambda$ is a conversion factor that accounts for all detector physics to produce units of flux from photon count.   
The GOES Databook lists the GOES long threshold flux ($F_{th} = 2 \times 10^{-8} W/m^{2}$) and sensitivity for a 10s exposure.
Therefore $\lambda$ is: $\lambda = \frac{10s\times F_{th}}{1} = 2\times10^{-7}$ Ws/m$^2$.
A measured flux is used to replace the value for counts and produce the error equation: $\sigma_{F} = \sqrt{\frac{F\lambda}{t_{exp}}}$
Peak flux measurements use a 1 min exposure time.  
Integrated flare curves calculate the error by using the summed flux over the total integration time $\sigma_{F_{total}} = \sqrt{\frac{F_{total}\lambda}{t_{total}}}$.
 
The error for equations used in calculations is determined through error propagation.
The error of a function X is represented as $\sigma_{X}$.
Known errors will be listed as $\delta_{Y}$ with the example: $Y \pm \delta_{Y}$. 

Error propagation for an equation of the form $M = A E^{\gamma}$ (such as equations (\ref{eq:Mcme}) and (\ref{eq:KEcme})) has the form:
\begin{equation}
\frac{\sigma_{M}}{M}=\sqrt{\left(\frac{\delta_{A}}{A}\right)^{2} + \left(\gamma\frac{\delta_{E}}{E}\right)^{2} + \left(Ln(E)\delta_{\gamma}\right)^{2} }
\end{equation}
.
M represents CME mass or CME kinetic energy based on which equation is used, E represents flare energy, $A$ is a constant of proportionality, and $\gamma$ in the power of which E is raised.
If the constant A is of the form $10^{\alpha \pm \delta\alpha}$, as it is for \citet{Drake2013}, then it becomes:
\begin{equation}
\frac{\sigma_{M}}{M}=\sqrt{\left(Ln(10)\delta_{\alpha}\right)^{2}+ \left(\gamma\frac{\delta_{E}}{E}\right)^{2} + \left(Ln(E)\delta_{\gamma}\right)^{2} }
\end{equation}
.
These terms are heavily dominated by the $Ln(E)\delta\gamma$ term and is the largest contributor to the error. 

Equation (\ref{eq:Combined}) has an error: 
\begin{equation}
\frac{\sigma_{E } } {E}=\frac{1}{1-\gamma}\sqrt{ \left(\frac{\delta_{A}}{A }\right)^{2} + \left(Ln\left[\frac{\epsilon Av^{2}}{2}f_{GOES}\right]\frac{\delta_{\gamma}}{1-\gamma}\right)^{2} + \left(2\frac{\delta_{v}}{v}\right)^{2} }
\end{equation} 
.
In the same manner as above, if A is of the form $10^{\alpha}$, then $\left(\frac{\delta_{A}}{A}\right)^{2}$ will become $\left(Ln(10)\delta_{\alpha}\right)^{2}$.
Here E represents flare energy, A and $\gamma$ are the constant of proportionality and exponent provided from equation (\ref{eq:Mcme}), $v$ is the velocity of the CME, and $f_{GOES} = 0.06$ is the proportion of flare energy within the GOES 1 - 8 \AA \ band \citep{Osten2015}. 

The error in equation (\ref{eq:VCombined}) is: 
\begin{equation}
\frac{\sigma_{v } } {v}=\frac{1}{2}\sqrt{ \left(\frac{\delta_{A}}{A }\right)^{2}  + \left((1-\gamma)\frac{\delta_{E}}{E}\right)^{2} + \left(Ln(E)\delta_{\gamma}\right)^{2}  }
\end{equation} 
.
If A is of the form $10^{\alpha}$, then $\left(\frac{\delta_{A}}{A}\right)^{2}$ will become $\left(Ln(10)\delta_{\alpha}\right)^{2}$.
Here $v$ is the velocity of the CME, A and $\gamma$ are the constant of proportionality and exponent  provided from equation (\ref{eq:Mcme}), E represents flare energy, and $f_{GOES} = 0.06$ is the proportion of flare energy within the GOES 1 - 8 \AA \ band \citep{Osten2015}. 

Finally, a linear fit model for a type II burst in plotted as Ln(Freq) vs Time has a measured slope (m).
The velocity is $\left(v = \frac{2H_{0}m}{cos\theta}\right)$.  
The error of the velocity given a slope m is:
\begin{equation}
\frac{\sigma_{v}}{v}=\sqrt{\left(\frac{\delta_{H_{0}}}{H_{0}}\right)^{2} + \left(\frac{\delta_{m}}{m}\right)^{2} + \left(tan(\theta)\delta_{\theta}\right)^{2}}
\end{equation}
where $v$ is the CME shock speed, $\theta$ is the angle the CME is oriented with respect to the radial direction, and $H_{0}$ is the density scale height of the corona. 
Using the assumption $\theta = 0 $ and $\delta_{\theta} = 0$, the error simplifies further.   

Rearranging equation (\ref{eq:equipar}) to solve for mass leads to: $M_{CME} = \frac{2E_{GOES}}{\epsilon f_{GOES}v^{2}}$.  
The error is:
\begin{equation}
\frac{\sigma_{M_{CME}}}{M_{CME}}=\sqrt{\left(\frac{\delta_{E_{GOES}}}{E_{GOES}}\right)^{2} + \left(\frac{2\delta_{v}}{v}\right)^{2}+ \left(\frac{\delta_{\epsilon}}{\epsilon}\right)^{2} + \left(\frac{\delta_{f_{GOES}}}{f_{GOES}}\right)^{2}}
\end{equation}
.

Any errors not covered in this section are straightforward and did not necessitate a unique comment.  

\section{Dynamic Spectra}

This section presents the individual events analyzed throughout this paper.

\subsection{X Class Events}

Figures 10 through 12 present GOES X-class events.  

\begin{figure}
\centering
\includegraphics[angle=-90,scale=.3,clip=true, trim=0cm 0cm 0cm 0cm]{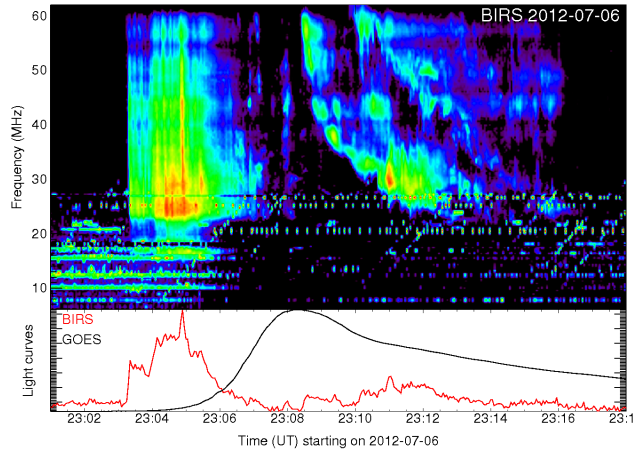} 
\caption{
Radio dynamic spectra and GOES light curve for X-class eruptive event occurring on 2012-07-06.
The top panel displays the BIRS radio dynamic spectra where a more red color denotes a more intense signal and a blue color is a fainter signal.
The bottom panel displays the GOES light curve in black.
It also displays a trace of a higher frequency line in the BIRS spectra range in red.  
Two events in the dynamic spectra; one exponential and one non-exponential.  
The exponential event (right most signal) was chosen for analysis. 
}
\end{figure} 

\begin{figure}
\centering
\includegraphics[angle=-90,scale=.3,clip=true, trim=0cm 0cm 0cm 0cm]{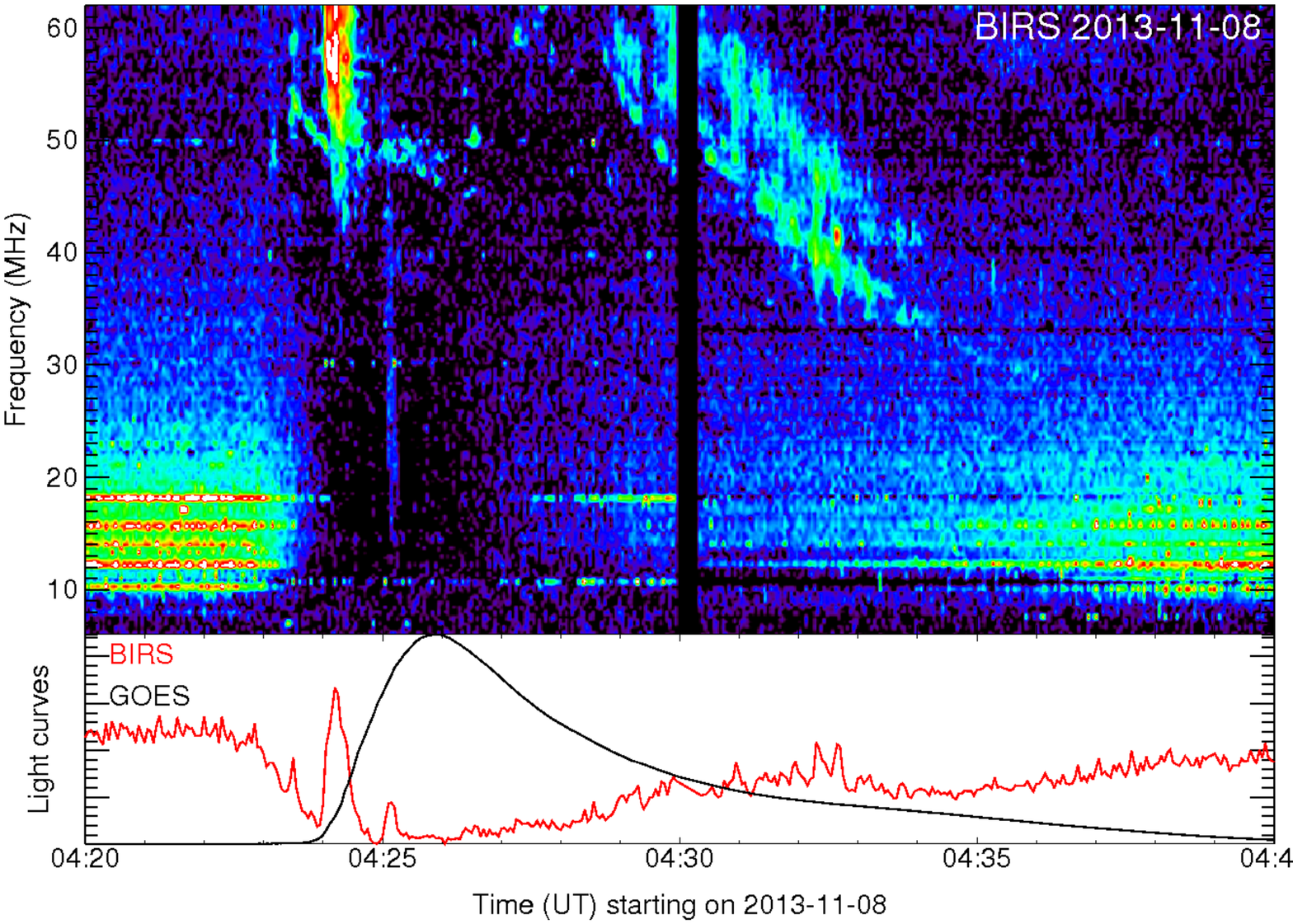} 
\caption{
Radio dynamic spectra and GOES light curve for X-class eruptive event occurring on 2013-11-08.
Figure format is the same as Figure 9.
}
\end{figure}

 \begin{figure}
\centering
\includegraphics[angle=-90,scale=.3,clip=true, trim=0cm 0cm 0cm 0cm]{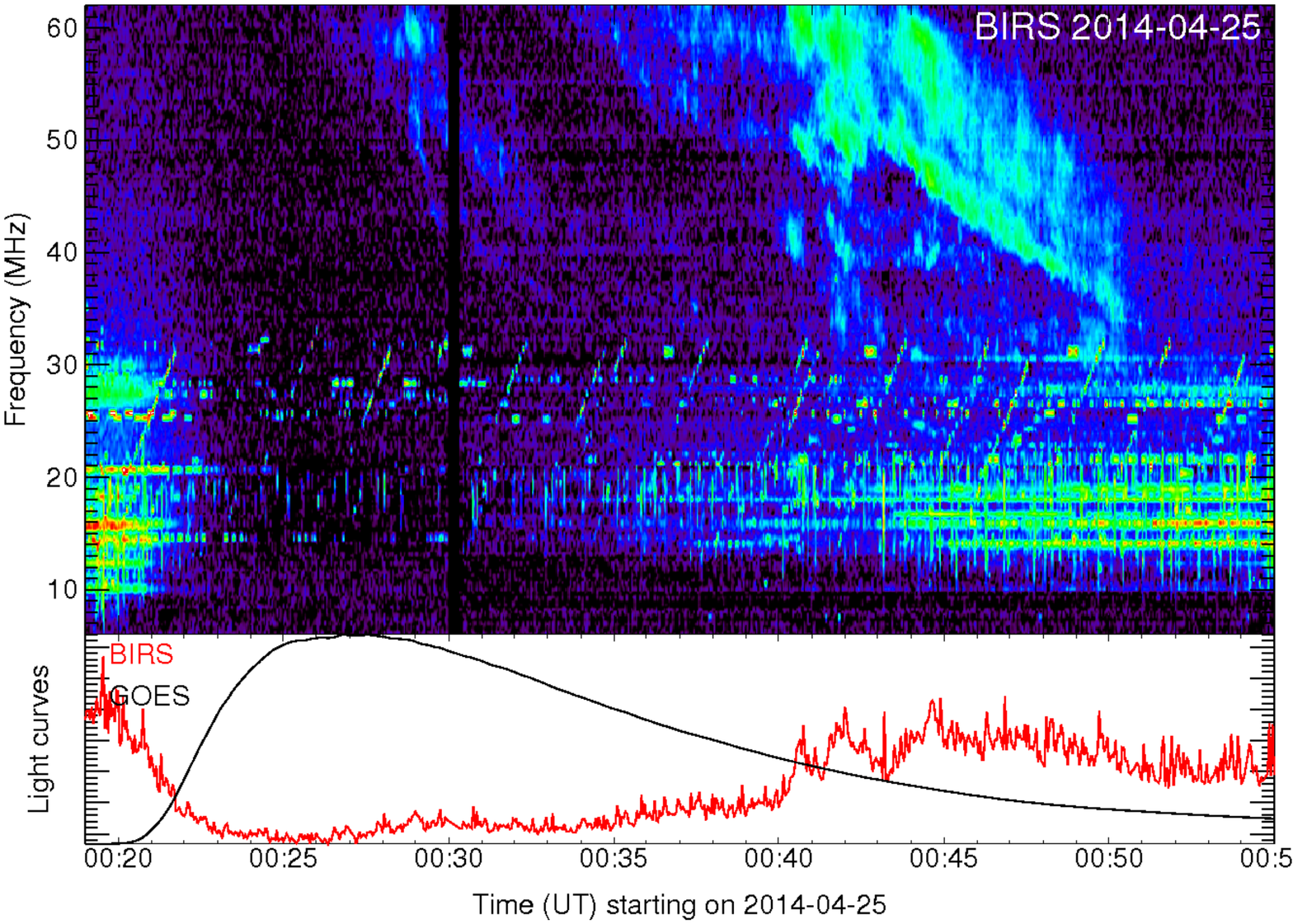} 
\caption{
Radio dynamic spectra and GOES light curve for X-class eruptive event occurring on 2014-04-25.
Figure format is the same as Figure 9.
}
\end{figure} 

\subsection{M Class Events}

Figures 13 through 22 present GOES M-class events.  

\begin{figure}
\centering
\includegraphics[angle=-90,scale=.3,clip=true, trim=0cm 0cm 0cm 0cm]{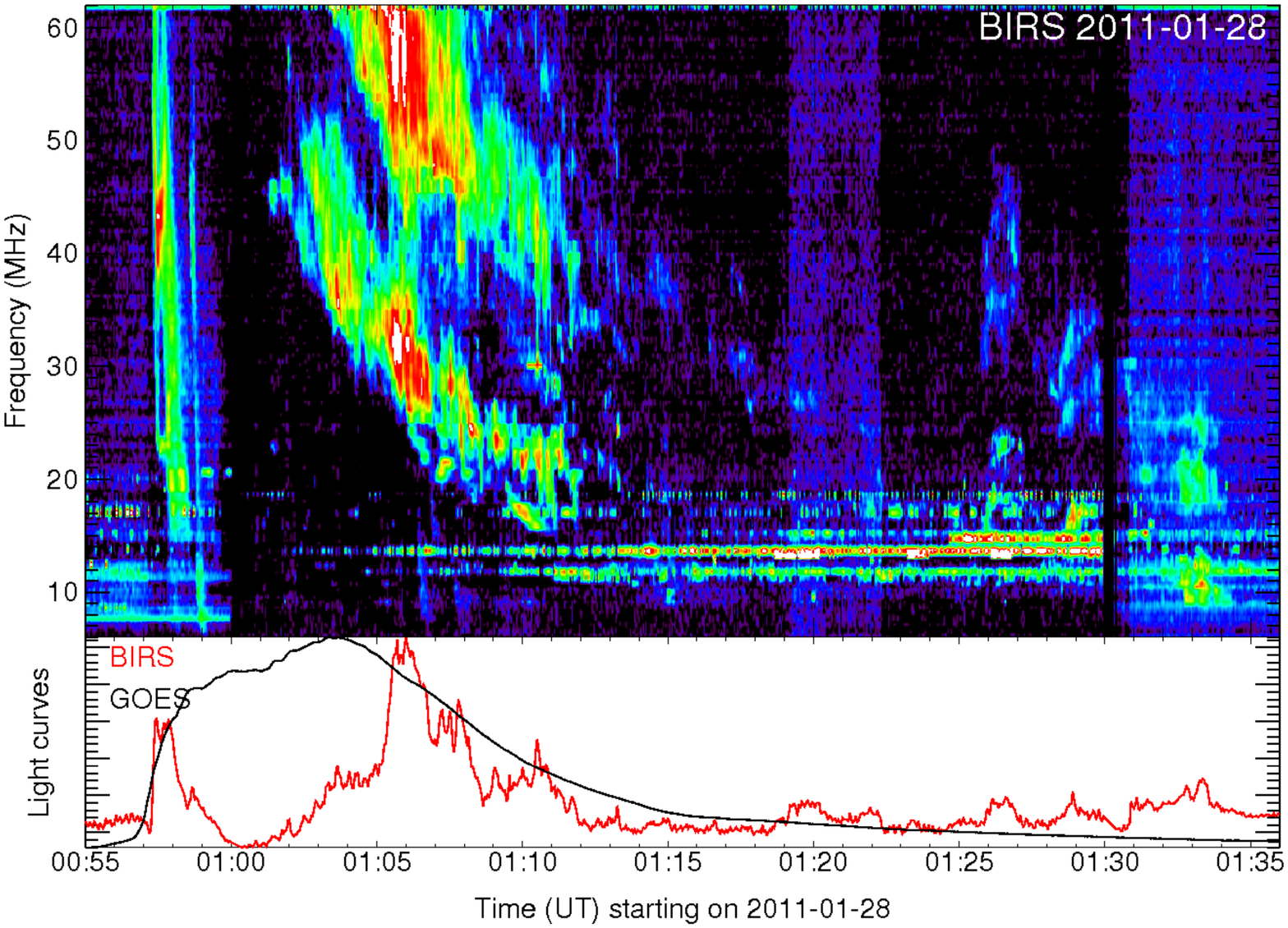} 
\caption{
Radio dynamic spectra and GOES light curve for M-class eruptive event occurring on 2011-01-28.
The top panel displays the BIRS radio dynamic spectra where a more red color denotes a more intense signal and a blue color is a fainter signal.
The bottom panel displays the GOES light curve in black.
It also displays a trace of a higher frequency line in the BIRS spectra range in red.  
}
\end{figure}

\begin{figure}
\centering
\includegraphics[angle=-90,scale=.3,clip=true, trim=0cm 0cm 0cm 0cm]{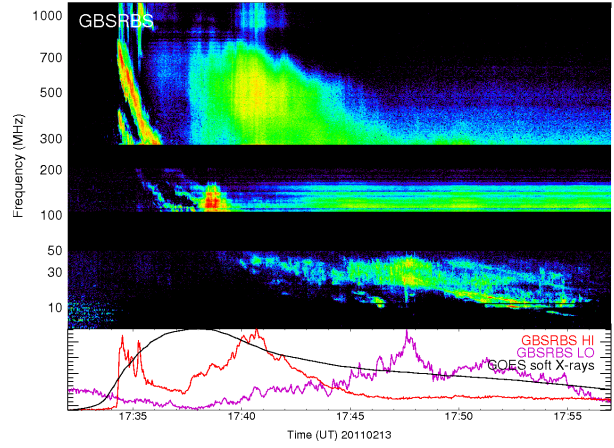} 
\caption{
Radio dynamic spectra and GOES light curve for M-class eruptive event occurring on 2011-02-13.
Figure format is the same as Figure 12.
CME is remarked as a Poor Event in the SOHO/LASCO catalog.
Event follows a non-exponential shape.  
}
\end{figure}

\begin{figure}
\centering
\includegraphics[angle=-90,scale=.3,clip=true, trim=0cm 0cm 0cm 0cm]{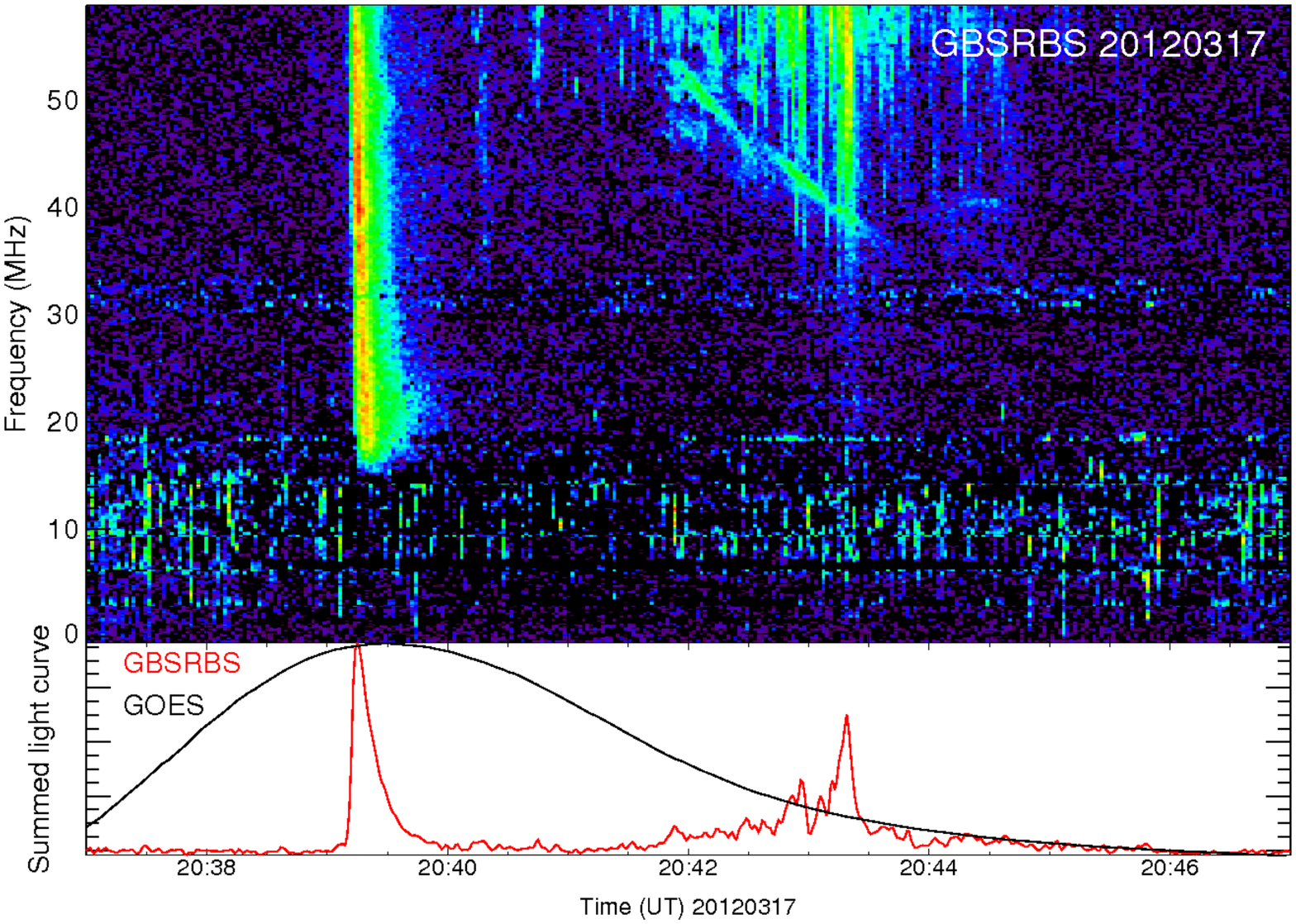} 
\caption{
Radio dynamic spectra and GOES light curve for M-class eruptive event occurring on 2012-03-17.
Figure format is the same as Figure 12.
CME remarked as a Poor Event in the SOHO/LASCO catalog.
}
\end{figure}

\begin{figure}
\centering
\includegraphics[angle=-90,scale=.3,clip=true, trim=0cm 0cm 0cm 0cm]{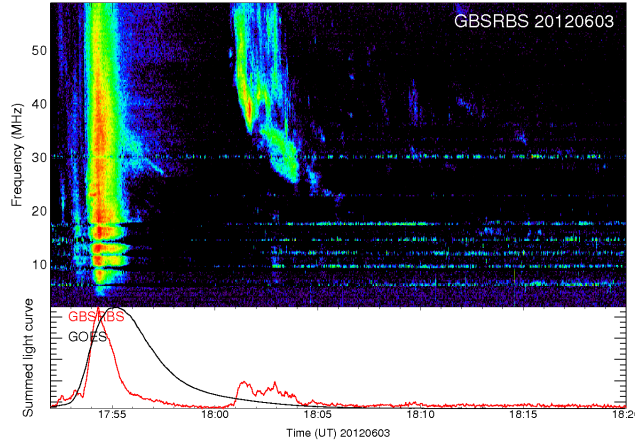} 
\caption{
Radio dynamic spectra and GOES light curve for M-class eruptive event occurring on 2012-06-03.
Figure format is the same as Figure 12.
Event follows a non-exponential shape.  
}
\end{figure}

\begin{figure}
\centering
\includegraphics[angle=-90,scale=.3,clip=true, trim=0cm 0cm 0cm 0cm]{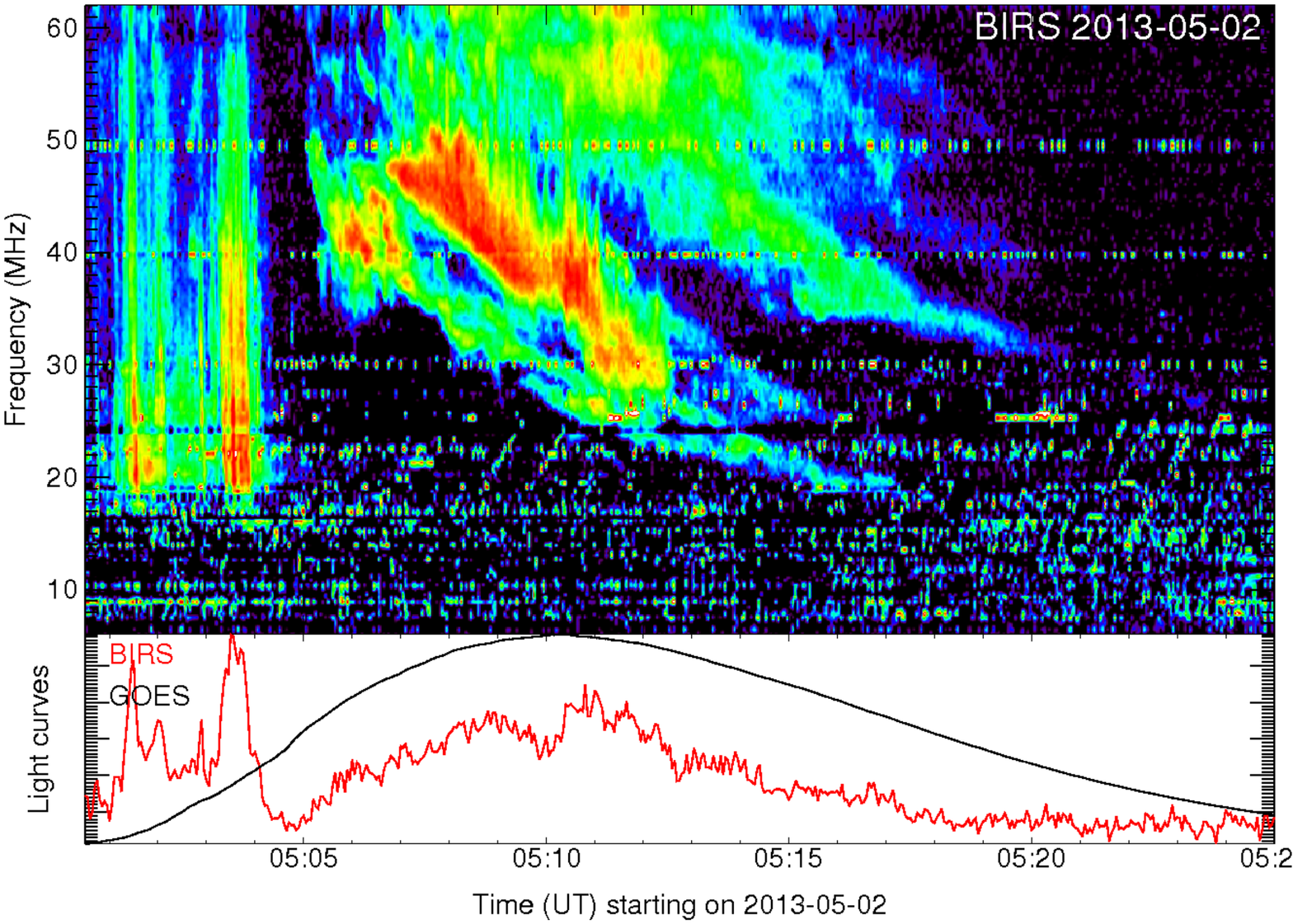} 
\caption{
Radio dynamic spectra and GOES light curve for M-class eruptive event occurring on 2013-05-02.
Figure format is the same as Figure 12.
}
\end{figure}

\begin{figure}
\centering
\includegraphics[angle=-90,scale=.3,clip=true, trim=0cm 0cm 0cm 0cm]{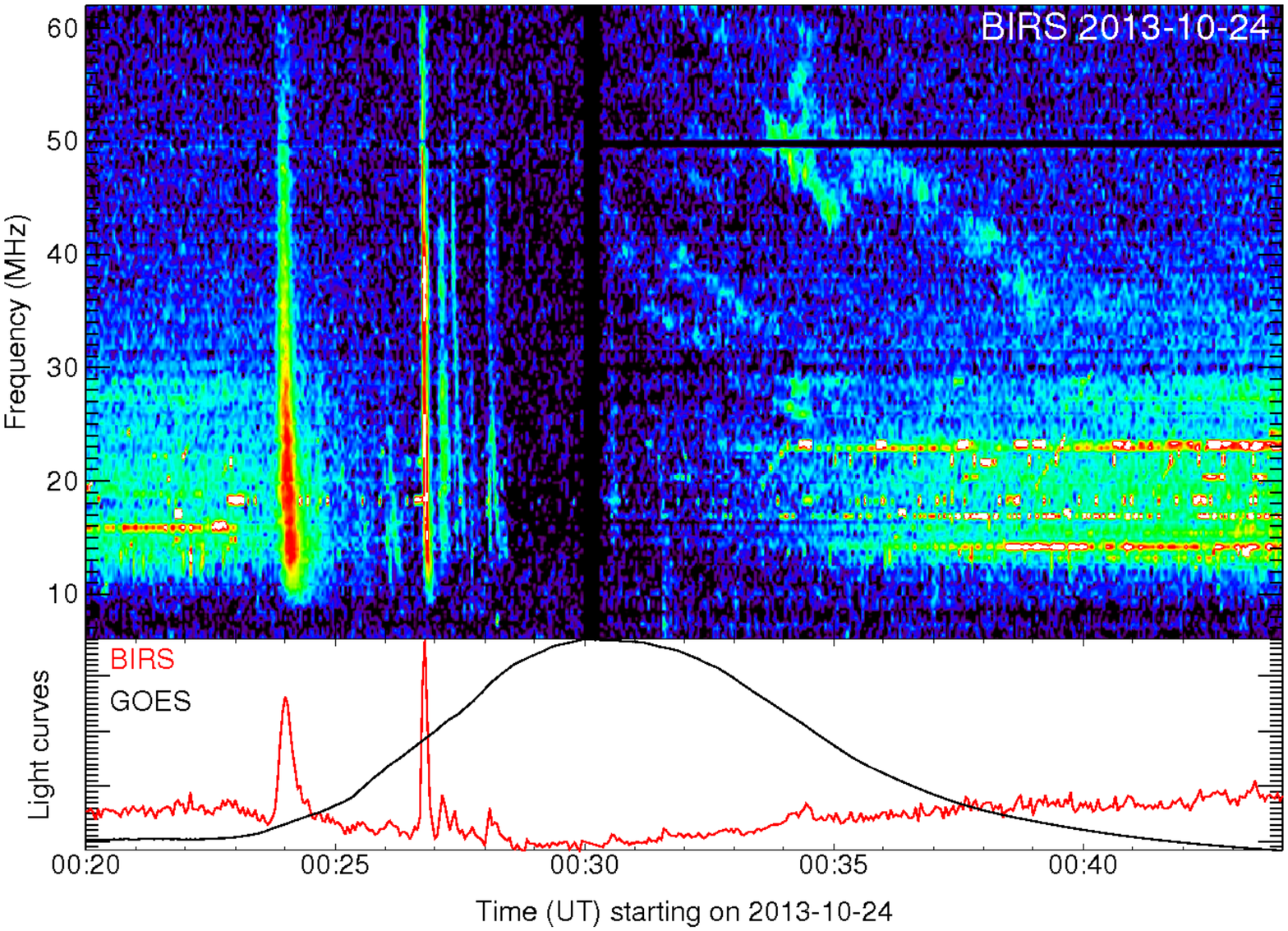} 
\caption{
Radio dynamic spectra and GOES light curve for M-class eruptive event occurring on 2013-10-24.
Figure format is the same as Figure 12.
CME remarked as a Poor Event in the SOHO/LASCO catalog.
}
\end{figure}

\begin{figure}
\centering
\includegraphics[angle=-90,scale=.3,clip=true, trim=0cm 0cm 0cm 0cm]{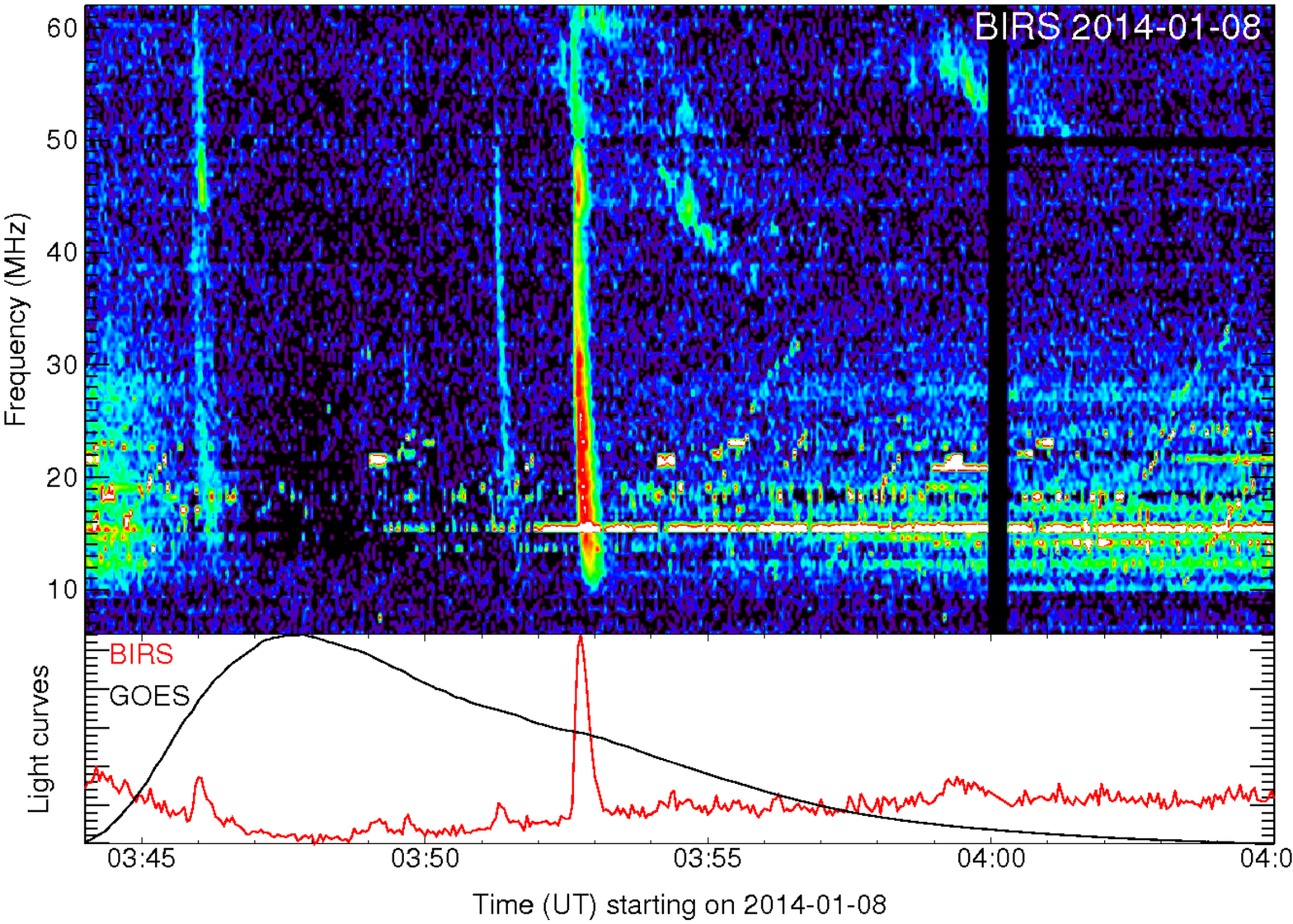} 
\caption{
Radio dynamic spectra and GOES light curve for M-class eruptive event occurring on 2014-01-08.
Figure format is the same as Figure 12.
CME remarked as a Poor Event in the SOHO/LASCO catalog.
}
\end{figure}

\begin{figure}
\centering
\includegraphics[angle=-90,scale=.3,clip=true, trim=0cm 0cm 0cm 0cm]{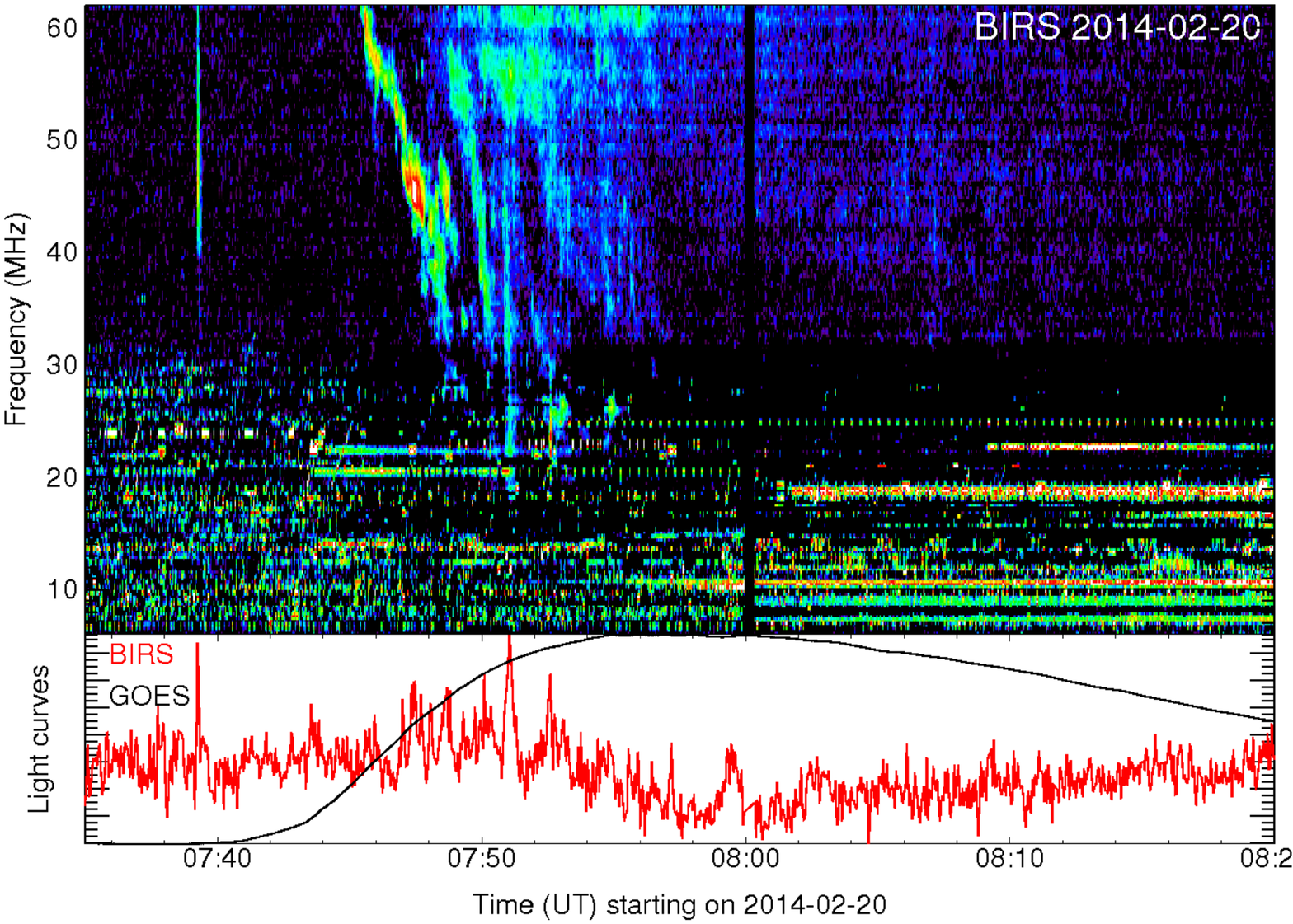} 
\caption{
Radio dynamic spectra and GOES light curve for M-class eruptive event occurring on 2014-02-20.
Figure format is the same as Figure 12.
CME remarked as a Poor Event in the SOHO/LASCO catalog.
}
\end{figure}

\begin{figure}
\centering
\includegraphics[angle=-90,scale=.3,clip=true, trim=0cm 0cm 0cm 0cm]{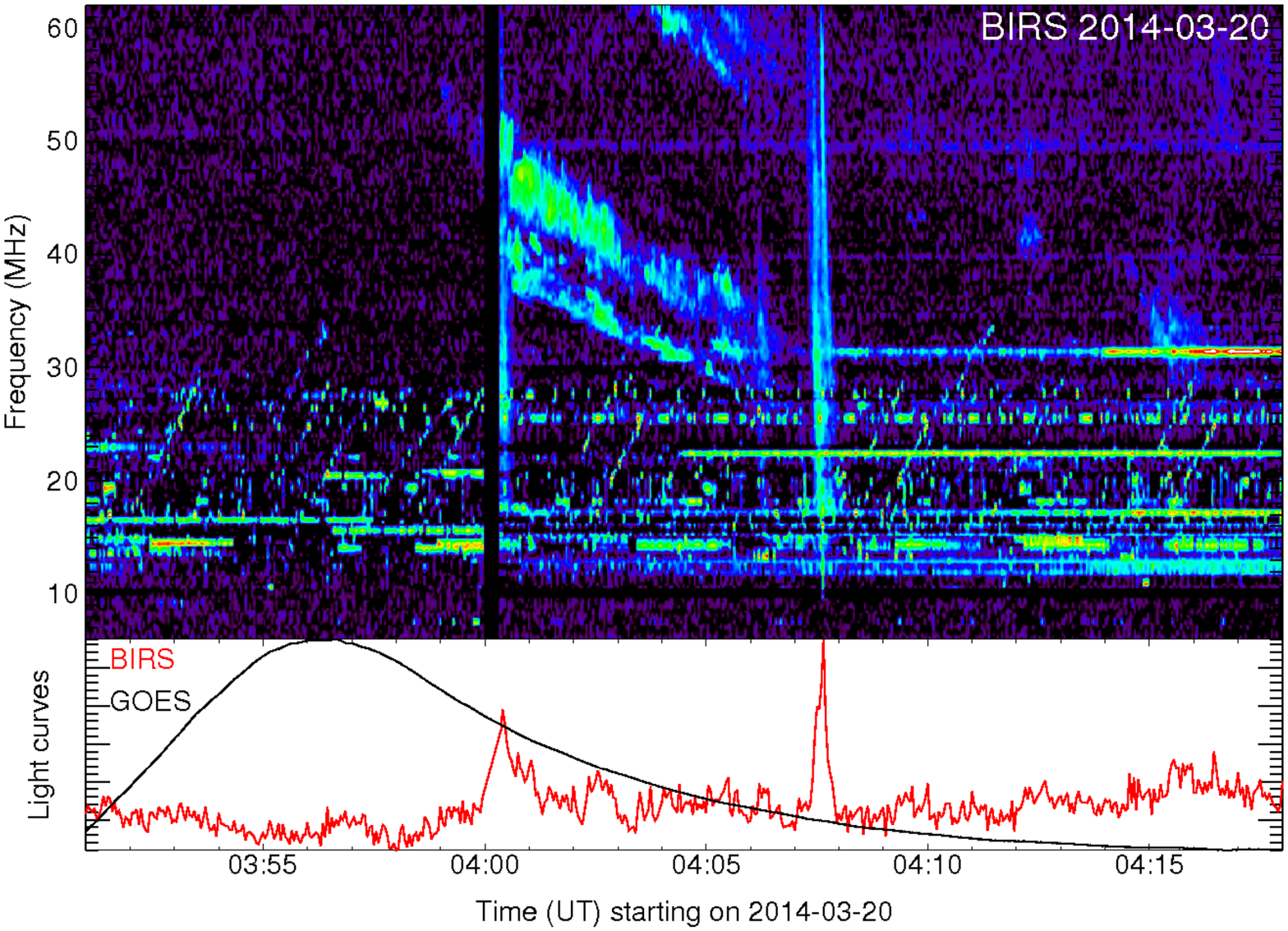} 
\caption{
Radio dynamic spectra and GOES light curve for M-class eruptive event occurring on 2014-03-20.
Figure format is the same as Figure 12.
}
\end{figure}

\begin{figure}
\centering
\includegraphics[angle=-90,scale=.3,clip=true, trim=0cm 0cm 0cm 0cm]{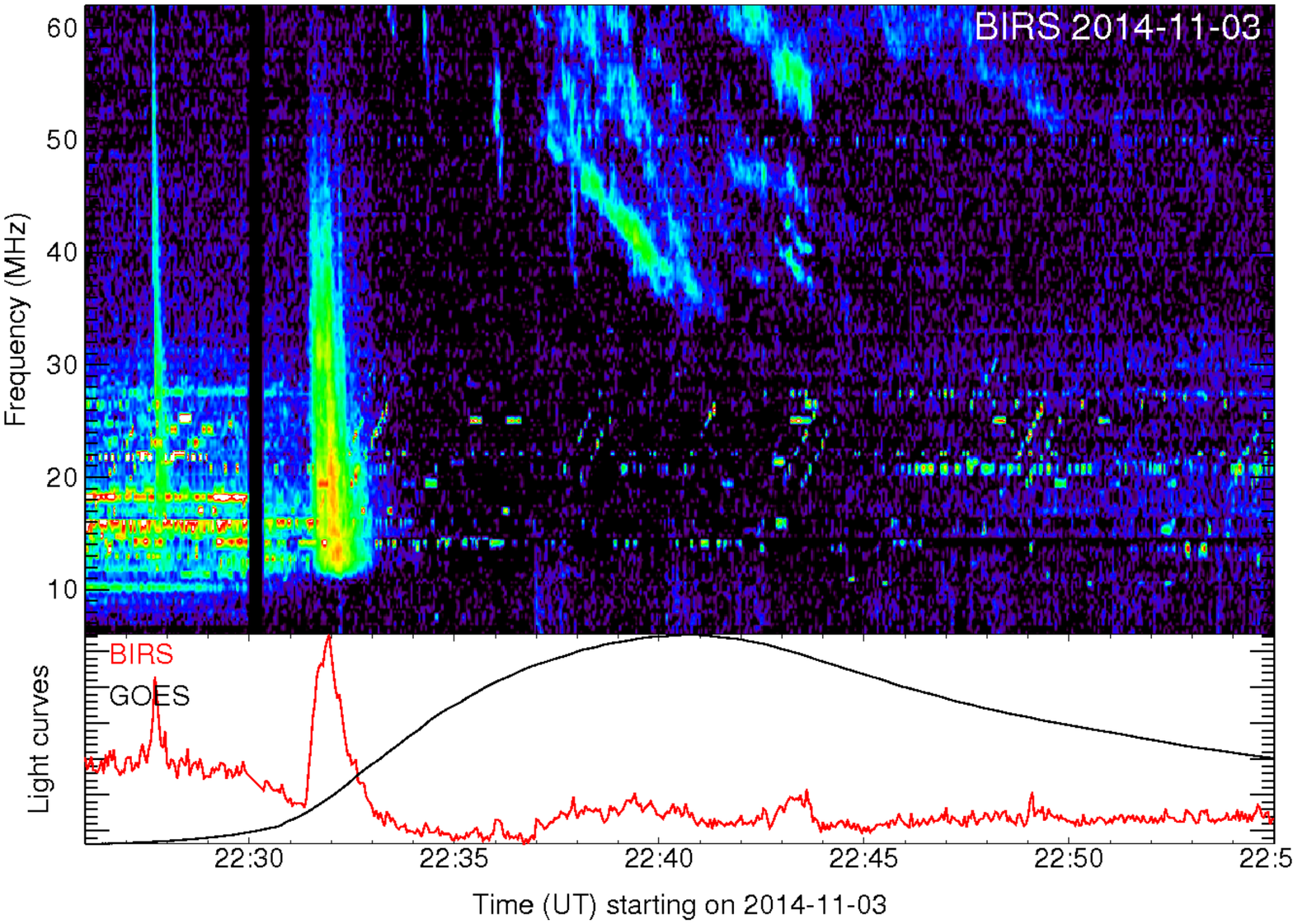} 
\caption{
Radio dynamic spectra and GOES light curve for M-class eruptive event occurring on 2014-11-03.
Figure format is the same as Figure 12.
}
\end{figure}

\begin{figure}
\centering
\includegraphics[angle=-90,scale=.3,clip=true, trim=0cm 0cm 0cm 0cm]{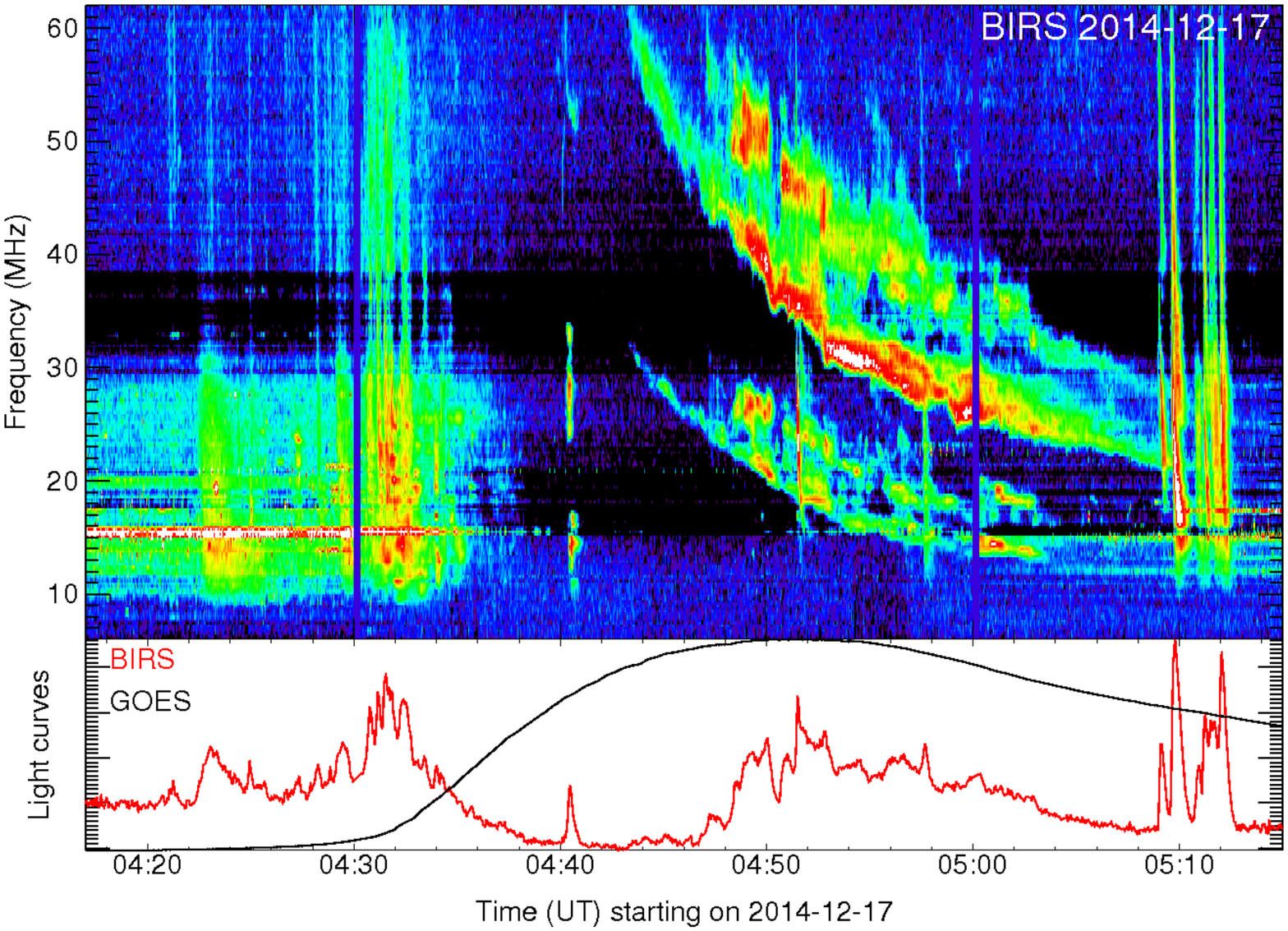} 
\caption{
Radio dynamic spectra and GOES light curve for M-class eruptive event occurring on 2014-12-17.
Figure format is the same as Figure 12.
}
\end{figure}

\end{document}